\documentstyle[preprint,aps]{revtex}

\newcommand{\sla}{\not\!}

\begin{document}

\font\fortssbx=cmssbx10 scaled \magstep2

\preprint{
\vbox{\hsize=5in\hbox{
\hskip.5in \raise.1in\hbox{\fortssbx University of Wisconsin - Madison}
}\hfill}
\vbox{
\hbox{\bf MAD/PH/800}
\hbox{December 1993}
\hbox{\hfil}}}

\title{Heavy Higgs Boson Production in Association with Three Jets at
Hadron Supercolliders}

\author{Adam Duff and Dieter Zeppenfeld}

\address{Department of Physics, University of Wisconsin, Madison, WI 53706 }

\maketitle

\begin{abstract}
   We consider the real emission QCD correction to heavy Higgs boson
production via weak boson fusion in high energy $pp$ collisions.  The
${\cal O}(\alpha_s)$ corrections are determined for the complete
electroweak $q q \rightarrow q q W^+ W^-$ process.  The presence of a
third parton in the final state affects the formation of rapidity gaps
only slightly.  In particular, soft emission into the gap region
is severely suppressed.  Also, we investigate how the additional hard
emission affects forward-jet-tagging and central-jet-vetoing efficiencies
in the search for $H \rightarrow W^+ W^- \rightarrow
\ell^+ \nu \ell^- \bar{\nu}$ decays.

\end{abstract}


\newpage

\section{Introduction}

One of the prime objectives of experiments at hadron supercolliders is
the discovery of the Higgs boson and the subsequent investigation of its
properties. If the Higgs boson is relatively heavy or if electroweak
symmetry breaking is driven by some new strong interactions, then the study
of Higgs boson production in the weak boson fusion process or of longitudinal
weak boson scattering becomes particularly important.

The study of the ``gold-plated'' $ZZ\to \ell^+\ell^- \ell^+\ell^-$ Higgs boson
decay mode alone will clearly not suffice to completely understand the nature
of electroweak symmetry breaking or may not yield a significant
signal~\cite{bagger}. Because of severe background problems one needs to
utilize as much information as possible about the full structure of the
underlying $qq\to qqVV$ process when trying to utilize other channels like
$W^+W^- \to \ell^+\nu \ell^-\bar{\nu}$ or hadronic decays of the
produced weak bosons.
This includes forward-jet-tagging~\cite{stirling,BCHOZ,BCHZ} to suppress
processes like $q\bar{q}\to VV$, central-jet-vetoing for the reduction
of top-quark backgrounds~\cite{BCHZ}, looking for relatively low overall
hadronic multiplicities~\cite{kane} or searching for rapidity gap
signatures~\cite{russians,bjgap} to exploit the color singlet exchange
between the two initial state quarks as a distinguishing feature of the
signal. In all these cases the knowledge of QCD emission corrections is
important or even crucial to assess the acceptance of the various
procedures for signal events.

In this paper we present first results of the calculation of all
${\cal O}(\alpha_s)$ real emission corrections to the full electroweak process
$qq\to qqWW$ (and crossing related ones) with subsequent leptonic $W$
decays. Results are obtained by the numerical evaluation of polarization
amplitudes~\cite{HELAS,HZ} and were implemented for $pp$ and $p\bar{p}$
collisions. An outline of the calculation is given in Section II, details
are relegated to an Appendix.

In two subsequent Sections we apply this calculation to two of the issues
mentioned above. The suppression of QCD radiation into the central rapidity
region is investigated in Section III. Color coherence between initial and
final state radiation off each of the incoming quarks and the absence of color
transfer between the two quark lines in the $qq\to qqWW$ process leads
to strongly suppressed radiation into the rapidity range between the two final
state quarks. This confirms the expectation that rapidity gaps, {\it i.e.}
regions of very low hadronic activity, may form between the two quark jets. The
influence of processes with gluons in the initial state ($e.g.$ $gq\to
q\bar{q}qWW$) on the radiation pattern and the effect of radiation on the
average gap width are analyzed here.

In Section IV we then investigate the effects of QCD radiation on
forward-jet-tagging and central-jet-vetoing as discussed in Ref.~\cite{BCHZ}.
Because of the additional jet activity the efficiency of a central-jet-veto
might be severely affected by higher order QCD corrections. We show that this
is indeed the case for the electroweak background of transverse $W$ production,
but not for the heavy Higgs signal itself.
We calculate the signal acceptance of the tagging and vetoing techniques
and find, typically, a 15\% reduction of signal rates compared to the lowest
order results.

\section{Calculational Methods}
\label{sec:2}

The processes which need to be considered are the QCD real emission
corrections to weak boson scattering. In the on-shell approximation for
the produced $W$'s this corresponds to the cross section evaluation at order
${\cal O}(\alpha_s \alpha_{QED}^4)$, {\it i.e.} at tree level, of the complete
process
\begin{equation}\label{processqqWWg}
q_1q_2 \to q_3q_4\; W^+W^-\; g
\end{equation}
and all crossing related processes, like, {\it e.g.},
\begin{equation}\label{processqqWWbarq}
gq_1 \to q_3\bar{q_2}q_4\; W^+W^-\; .
\end{equation}
Some representative Feynman graphs are shown in Fig.~\ref{figone} and the full
set of contributing graphs is outlined in the Appendix. We are primarily
interested in the weak boson scattering process as depicted in Figs.~1a and b.
However, when trying to calculate the distributions of the final state quarks
we need to treat the incoming weak bosons as off-shell particles.
Electromagnetic gauge
invariance then requires to consider the $t$-channel photon exchange of
Fig.~1b together with $W$ bremsstrahlung off the two quark lines. We are thus
lead to consider the full ${\cal O}(\alpha_s \alpha_{QED}^4)$ process,
including all $W$ bremsstrahlung processes off the external quark lines,
as shown, for example, in Fig.~1c.

Questions of electromagnetic gauge invariance do not arise when considering
the $s$-channel Higgs resonance only, as depicted in Fig.~1a. For a simplified
definition of the Higgs boson signal
we shall also consider these ``$s$-channel resonance''
contributions in isolation. It is well known, however, that for a large Higgs
boson width, {\it i.e.} in the case of a heavy Higgs boson, the resonance
contribution overestimates the production cross section of longitudinal $W$'s
at large values of the $W$ pair invariant mass~\cite{dicus}. The ``$s$-channel
resonance'' contributions should therefore only be taken as a qualitative
estimate of the Higgs boson signal. As an alternative means of
isolating the effects of a heavy Higgs boson or of the scattering of
longitudinal weak bosons we shall also consider the cross section difference
$\sigma(m_H) - \sigma(m_H=100\, {\rm GeV})$, {\it i.e.} the excess events
over a light Higgs boson scenario.

The lowest order process, without the gluon emission considered here, has
been calculated by several groups~\cite{dicus,BCHZ}. We obtain the real
emission QCD corrections by numerically evaluating the amplitudes, using the
helicity amplitude calculus of Ref.~\cite{HZ} as implemented in the HELAS
package~\cite{HELAS}. Fermion masses are neglected everywhere. When
considering leptonic $W$-decays the $W$ propagator factors are taken in
the narrow width approximation in order to eliminate additional Feynman
graphs which are required by gauge invariance for off-shell $W$'s. Details
are given in the Appendix.

A simple Breit-Wigner resonance form with a constant width in the timelike
region is used for the
Higgs boson propagator. Refinements can be made~\cite{Will90}, but they are not
relevant in the following since we are primarily interested here in the QCD
structure of weak boson fusion processes and do not aim at a precise modeling
of the SM prediction for $W^+W^-+3$ jet production. Finally, the phase space
integrals over the squared amplitudes are performed with the VEGAS integration
package~\cite{vegas}. For the parton densities inside the proton we use set
$D_-'$ of Ref.~\cite{MRSD} and we choose the geometric mean of the final state
parton transverse momenta as the scale of the structure functions and of the
strong coupling constant $\alpha_s(Q^2)$.

Because we are working in the Born approximation and since we are using
massless quarks throughout, the total cross section for $pp \to W^+W^-X$
comes out to be divergent and we need to impose acceptance cuts to obtain
finite results.

$t$-channel photon exchange as shown in Fig.~1c is singular at low $Q^2$ and
in addition the parton model no longer provides an adequate description in
this phase space region. Rather, one should use measured electromagnetic form
factors of the proton. While a proper treatment of low $Q^2$ photon exchange
is possible~\cite{BVZ} one finds that this region can safely be neglected
when considering signatures for weak boson scattering~\cite{BCHOZ}.
Hence, we only consider the deep inelastic scattering region in the following
by imposing the cut $|Q^2| > 4\; {\rm GeV}^2$.

Annihilation diagrams contribute in processes with identical flavors on the
two quark lines. An example is shown in Fig.~1d. Here the splitting of the
$s$-channel photon into a $q\bar{q}$ pair diverges at small dijet invariant
masses. Again this process is unimportant for the weak boson scattering
regime in which we are interested. We eliminate the singularity by requiring
$m_{jj} > 10\; {\rm GeV}$ for all pairs of final state partons. Finally,
collinear singularities, which arise from initial state gluon radiation, are
avoided by considering partons of finite transverse momentum only.

These cuts eliminate all numerical divergencies. They are largely superseded by
the acceptance cuts which will be made in actual experiments. Throughout the
following we require that all pairs of partons which satisfy the
jet-identification requirements (typically $p_T > 40$~GeV and a limited
range in pseudorapidity)
are well separated in the pseudorapidity--azimuthal angle plane,
\begin{equation}\label{Rjj}
R_{jj} = (\Delta\eta_{jj}^2+\Delta\phi_{jj}^2)^{1\over 2} > 0.7\; ,
\end{equation}
and similarly that all decay leptons are isolated,
\begin{equation}
R_{\ell j} = (\Delta\eta_{\ell j}^2+\Delta\phi_{\ell j}^2)^{1\over 2} > 0.7\; .
\end{equation}
Additional requirements on jet and lepton transverse momenta and on the
angular acceptance will be listed separately in the following sections.

We will want to study the effects of the ${\cal O}(\alpha_s)$
QCD corrections on one and two jet inclusive distributions and we would like
to use the code to assess the probability of soft radiation into the rapidity
interval defined by two tagging jets. A complete calculation of such
observables would require the determination of virtual effects and the
resummation of soft gluon emission, tasks which are clearly beyond the
scope of the present paper. Instead we use the ``truncated shower
approximation'' (TSA) to obtain estimates of these observables~\cite{pps}.

In the TSA we replace the tree level differential cross section for three
parton final states, $d\sigma(WWjjj)_{\rm TL}$, by
\begin{equation}
d\sigma(WWjjj)_{\rm TSA}=d\sigma(WWjjj)_{\rm TL}
     \left(1-e^{- p_{Tj,min}^2/p_{tsa}^2}\right)\;, \label{reg}
\end{equation}
Here $p_{Tj,min}$ is the smallest transverse momentum of the three final
state partons. As $p_{Tj}\to 0$ the final factor in Eq.~(\ref{reg}) acts as
a regulator of the small $p_T$ singularity. The TSA parameter $p_{tsa}$ is
chosen to reproduce the lowest order cross section within a given set of
acceptance cuts.
For the full electroweak process the singularities associated with low $Q^2$
 photon exchange would introduce a very strong dependence of $p_{tsa}$ on the
phase space region which is considered. This is not the case, however, for
the Higgs boson signal as defined by the ``$s$-channel resonance''
contribution. It is well known that the $K$-factor for the $qq\to qqH$
process is close to unity~\cite{hvw} and this process has a finite total
cross section at tree level. By choosing $p_{tsa}$ such that the ``$s$-channel
resonance'' contribution to $\sigma(WWjjj)_{\rm TSA}$  reproduces the
corresponding lowest order cross section, we obtain an algorithm which
agrees with the full ${\cal O}(\alpha_s \alpha_{QED}^4)$ calculation in
the 3 jet phase space region, which allows to study 1-jet and 2-jet
inclusive distributions to the same order, and which provides an excellent
approximation to these distributions and to the overall normalization at
full  ${\cal O}(\alpha_s \alpha_{QED}^4)$ in the phase space region where
the Higgs resonance dominates.

Anticipating the acceptance cuts to be used below, we have determined the
TSA parameter by matching the total Higgs production cross sections in
$pp$ scattering at $\sqrt{s}=40$~TeV with one or two visible jets exceeding
a minimal transverse momentum. Results are given in Table I. One finds that
$p_{tsa}$ is quite insensitive to the value of the Higgs boson mass and also
varies very little with the $p_{Tj,min}$ cut imposed in the 1-jet inclusive
case. When requiring the presence of two partons of $p_T>$ 20, 40, or 60 GeV
this variation is somewhat stronger, but even here the variation in $p_{tsa}$
of $\approx\pm 1$~GeV corresponds to normalization changes of the cross
section by up to 30\% only, which is within the uncertainty range of our
tree level calculation.

\section{Color Structure and Rapidity Gaps}
\label{sec:3}

The characteristic features of the ${\cal O}(\alpha_s)$ emission corrections
to weak boson scattering can be understood in terms of the color
structure of the lowest order process. In the dominant $t$-channel
contributions (see Figs.~1a--c) no color is exchanged between the two
incoming quarks (or anti-quarks). For small scattering angles of these
incoming fermions the color charges are accelerated rather little and the
resulting ``synchrotron radiation'' of gluons occurs predominantly into the
forward direction, between the beam axis and the direction of the scattered
(anti)quark. One thus expects very little hadronic activity to arise from
bremsstrahlung off the hard process in the central region, between the two
scattered quarks. Since the energy carried by the quarks is typically much
larger than the virtuality of the emitted weak bosons (of ${\cal O}(m_W)$)
these quarks are produced at rather large pseudorapidities, leaving a wide
rapidity region with suppressed radiation~\cite{russians}. This pattern is
fundamentally different from the one expected for typical background
processes (like $e.g.\; gg\to t\bar{t}\to b\bar{b}W^+W^-$) where color
exchange between the two incoming partons leads to strong gluon radiation
into all regions of the legoplot.

The suppressed radiation into the central region may then lead to the
formation of rapidity gaps, pseudorapidity regions of very low or no
hadronic activity except for the Higgs decay products~\cite{russians,bjgap}.
This rapidity gap may be filled, however, by hadrons from the underlying
event~\cite{sjo} and the full formation of the gap is expected to be
observable only in a small fraction of all signal events where fluctuations
or the absence of multiple parton interactions make the radiation pattern of
the hard scattering process visible~\cite{bjgap,fletcher,gotsman}.

Here we do not discuss this last question of the gap survival probability
any further, rather we investigate whether the higher order QCD corrections
indeed do lead to the radiation patterns suggested by the color flow
arguments. Processes with three quarks in the final state
(like $.e.g$ $g q \to q\bar{q}qH$) might substantially alter the expected
pattern. In addition, a precise knowledge of the radiation in signal
vs. background events may help to distinguish the two even
in the presence of an underlying event~\cite{kane}.

The features of the heavy Higgs boson signal discussed above suggest a search
following the strategy developed for rapidity gap events in Ref.~\cite{CZ}.
We study 2 jet inclusive $WW$ events where both $W$'s decay leptonically and
where the two charged decay leptons fall into the central rapidity region
between the two tagging jets. More precisely we start with events containing
two charged leptons of high transverse momentum in the central region,
\begin{equation}
\label{cutell}
p_{T\ell} > 100\; {\rm GeV},\qquad   |\eta_\ell| < 2\; .
\end{equation}
No hadronic jet with transverse momentum
\begin{equation}\label{ptjet}
p_{Tj} > 40\; {\rm GeV}
\end{equation}
is allowed in the pseudorapidity interval between the two leptons. On either
side of the lepton pair we search for the first jet of $p_{Tj}>40$~GeV, $i.e.$
we require the presence of two tagging jets. The pseudorapidity range between
the tangents to the jet definition cones (of radius $R=0.7$) is called the
``gap region'' and the two charged leptons are required to fall into this gap
region. Denoting by $\eta_{j_1} < \eta_{j_2}$ the pseudorapidities of the two
tagging jets we thus require
\begin{equation}\label{defgap}
\eta_{j_1}+0.7 < \eta_{\ell^+},\, \eta_{\ell^-} < \eta_{j_2}-0.7 \; .
\end{equation}

The width of the gap region is denoted by
\begin{equation}\label{defygap}
y_{gap} = |\eta_{j_1} - \eta_{j_2}|-2\cdot 0.7 \; .
\end{equation}
The average gap width is expected to be large for weak boson scattering events
but not necessarily in the phase space region where $W$ bremsstrahlung
dominates. This is confirmed in Fig.~\ref{ygap} where $d\sigma /dy_{gap}$,
within the acceptance requirements of Eqs.~(\ref{cutell})--(\ref{defgap}),
is shown
for the production of a $m_H = 800$~GeV Higgs boson in $pp$ collisions at
$\sqrt{s}=40$~TeV. The Higgs boson and longitudinal weak boson scattering
contributions can be isolated by comparing with the case of a very light Higgs
boson, taken here as $m_H=100$~GeV. For the lepton acceptance requirements of
Eq.~(\ref{cutell}) the resonance contribution of the light Higgs boson is
completely negligible. Thus the $m_H=100$~GeV curve represents an estimate
of the contributions due to $W$ bremsstrahlung and transverse $W$ production,
which may be called an ``electroweak background'' to the heavy Higgs boson
signal. Comparison of the light and heavy Higgs boson curves in Fig.~\ref{ygap}
indeed shows that the signal events are produced at large average $y_{gap}$.

The main effect of the third parton in the final state is a slight narrowing of
the average gap width from $<y_{gap}>=4.9$ in the lowest order calculation to
$<y_{gap}>=4.5$ in the full ${\cal O}(\alpha_s)$ result (for $m_H=800$~GeV).
There are at least two reasons for this effect: the possible emission of
a third jet into the region between the original two quark jets redefines the
gap region and in addition the emission of a third parton outside the gap
region, $i.e.$ at very large rapidities, raises the center of mass energy of
the event and hence leads to a kinematical reduction of the gap width. Neither
effect is very large, however.

The QCD radiation of the third parton is dominated by collinear emission
close to one of the two tagging jets. Because of the coherence between initial
and final state radiation the emission occurs mainly between the tagging jets
and the beam axes, $i.e.$ outside the gap region. This effect is demonstrated
in Fig.~\ref{ysclosest} where the distribution in the rapidity difference of
the third (typically soft) parton and the closest tagging jet is shown,
\begin{equation}
\Delta\eta_{sj} = {\rm sign}\cdot (\eta_{\rm soft}-\eta_{j,\rm closest})\; .
\end{equation}
The sign is chosen such that positive $\Delta\eta_{sj}$ corresponds to
radiation outside the interval marked by the two tagging jets and negative
values correspond to
radiation towards the gap region. The edge of the gap region is visible as a
step at $\Delta\eta_{sj}=-0.7$ which is a result of the jet separation
requirement of Eq.~(\ref{Rjj}). Clearly emission of the third parton outside
the gap region is preferred. This is true both for the heavy Higgs case (solid
line) and for the electroweak transverse $W$ background, simulated by the
$m_H=100$~GeV scenario (dashed line).

The resulting radiation pattern is best appreciated by choosing the center of
the gap region as the origin and then normalizing the pseudorapidity of the
third parton to the gap width. This is achieved by using the
variable~\cite{duff}
\begin{equation}\label{defofz}
z = {2\eta_{\rm soft} - \eta_{j_1} -\eta_{j_2} \over
     |\eta_{j_1}-\eta_{j_2}|-1.4}\; .
\end{equation}
Thus $z=0$ corresponds to to the gap center while $z=\pm 1$ indicates the
edges of the gap region. The resulting radiation pattern, $d\sigma/ dz$, is
shown in Fig.~\ref{dsigdz}. The probability for radiation into the gap region
is strongly suppressed both for the heavy Higgs signal and for the electroweak
background. The radiation pattern which is expected for $t$-channel color
singlet exchange is thus confirmed at ${\cal O}(\alpha_s)$.

\section{QCD Effects on Jet-Tagging and Central-jet-vetoing}
\label{sec:4}

When trying to observe the $W^+W^-$ decay mode of a heavy Higgs boson at a
hadron collider, one needs to fight serious physics backgrounds. The production
of a top-quark pair with subsequent decay $t\bar{t}\to bW^+\bar{b}W^-$ is
particularly troublesome due to the large rate of this background. Of
somewhat lesser importance is $q\bar{q}\to W^+W^-$ pair production~\cite{BCHZ}.
Higgs boson production via weak boson fusion differs in important aspects
from these background processes. Forward-jet-tagging of the quark-jet(s) off
which the initial state $W$'s or $Z$'s were radiated offers a powerful tool for
background suppression~\cite{stirling}. In addition one can use the fact that
the $W$'s arising from top-quark decay are always accompanied by a nearby
$b$-quark which often manifests itself as an additional hadronic jet. When
searching for the leptonic decays of the produced $W$'s in the central rapidity
region a veto on any additional central jet constitutes a powerful tool for
top-background reduction.

Single forward-jet-tagging and central-jet-vetoing were analysed in
Ref.~\cite{BCHZ}. The Higgs boson signal was simulated with a full
${\cal O}(\alpha_{\rm QED}^4)$ Monte Carlo program. While this appears adequate
for the analysis of tagging jet distributions, the acceptance of signal
events under severe central-jet-vetoing conditions might be strongly degraded
by ${\cal O}(\alpha_s)$ QCD corrections. This question can be addressed with
the tools described in Section II.

Following Ref.~\cite{BCHZ} we consider purely leptonic decays of the two $W$
bosons and concentrate on the Higgs resonance region by imposing stringent
lepton acceptance cuts,
\begin{equation}\label{cutell2}
|\eta_\ell| < 2\;,\qquad  p_{T\ell} > 100\; {\rm GeV},\qquad
\Delta p_{T\ell\ell} > 400 {\rm GeV}\;  .
\end{equation}
Here $\Delta p_{T\ell\ell} = |{\bf p}_{T\ell_1}-{\bf p}_{T\ell_2}|$ is the
difference of the two charged lepton transverse momenta~\cite{DGV}.

In a second step one requires the presence of at least one hadronic jet of
transverse momentum
\begin{equation}\label{ptjettag}
p_{Tj} > 40\; {\rm GeV}\; .
\end{equation}
The most energetic such jet is the tagging jet candidate. In Figs.~\ref{ytag}
and \ref{pttag} the pseudorapidity and transverse momentum distributions of
this jet, as obtained with the full ${\cal O}(\alpha_s\alpha_{\rm QED}^4)$
Monte Carlo,
are shown for a $m_H=800$~GeV signal and the electroweak background as defined
by a $m_H=100$~GeV scenario. Most of the tagging jet candidates at low
$\eta_{j,\rm tag}$ arise from transverse $W$ bremsstrahlung
and hence a stringent
rapidity cut on the tagging jet is highly efficient for the heavy Higgs signal.
The transverse momentum of the tagging jet in Higgs signal events
is typically quite low, while
the electroweak background distribution is rather flat, reflecting the
dominance of $W$ bremsstrahlung: the strong lepton $p_T$ cut selects high $p_T$
$W$-bosons which were radiated off high transverse momentum quarks. The
tagging jet distributions are quite similar to the
${\cal O}(\alpha_{\rm QED}^4)$ calculation which, therefore, are not shown
separately. As in Ref.~\cite{BCHZ} signal events are selected by requiring the
presence of a tagging jet of
\begin{equation}\label{Ejtag}
E_{j,\rm tag} > 1\; {\rm TeV}\; ,\qquad   3<|\eta_{j,\rm tag}|<5\; .
\end{equation}

Because of the presence of a third colored parton in the final state the full
${\cal O}(\alpha_s\alpha_{\rm QED}^4)$ calculation yields a larger probability
to find an additional jet in the central region than the corresponding result
without QCD corrections. Let us define the veto jet candidate as the remaining
parton, other than the tagging jet, with largest transverse momentum.
The pseudorapidity distribution of this veto jet candidate is shown in
Fig.~\ref{etaveto}.

In Ref.~\cite{BCHZ} a veto of any jet in the central region satisfying the
conditions
\begin{equation}\label{cutveto}
|\eta_j({\rm veto})| < 3\;,\qquad  p_T({\rm veto}) > 30\; {\rm GeV}\;
\end{equation}
was found to provide adequate top background rejection while retaining most of
the signal events when neglecting QCD corrections to the Higgs signal. The
cross sections for the Higgs boson signal after vetoing central jets with
varying minimum transverse momentum requirements are shown in Table II for both
the lowest order ``2 parton calculation'' and the ${\cal O}(\alpha_s)$ ``3
parton calculation''. Also given are the efficiencies of retaining the Higgs
signal after central-jet-vetoing. The 100\% level corresponds to the cross
section difference $\sigma(m_H=800\,{\rm GeV})-\sigma(m_H=100\,{\rm GeV})$
before rejecting events with jets of $|\eta_j|<3$ and transverse momenta as
listed in the table.

In the phase space region which is dominated by $W$ bremsstrahlung (as
simulated by the $m_H=100$~GeV scenario) QCD radiation does indeed
strongly affect the efficiency of a central-jet-veto. In fact this leads to a
further reduction of the electroweak background compared to the Higgs boson
signal. At ${\cal O}(\alpha_s)$ the signal
acceptance of the central-jet-veto is lowered by $\approx 20\%$ and the
resulting signal cross sections are typically only $\approx 15\%$ lower
than before QCD corrections. Hence the conclusions of Ref.~\cite{BCHZ}
remain valid when taking real emission QCD corrections into account.

\bigskip
{\bf Acknowledgements}
This research was supported in part by the University of Wisconsin Research
Committee with funds granted by the Wisconsin Alumni Research Foundation,
by the U.~S.~Department of Energy under contract No.~DE-AC02-76ER00881,
and by the Texas National Research Laboratory Commission under Grants
No.~RGFY9273 and FCFY9212.

\newpage
\appendix
\section{Matrix Elements}
\label{app:1}

The external state of the processes considered in Section~\ref{sec:2}
consists of four quarks, one gluon, a $W^+$ and a $W^-$ boson.  The $W^\pm$
bosons will be taken later to decay into leptons, but this does
not change the nature of the calculation.  The four (anti)quarks
are labelled as $a, b, c, d$, along with their corresponding momenta
$p_a, p_b, p_c, p_d$, and helicities $\lambda_a, \lambda_b, \lambda_c,
\lambda_d$.  The four (anti)quarks are paired to
form two fermion lines, denoted as $(ac)$ for fermion flow from
$a \rightarrow c$, and $(bd)$ for fermion flow from $b \rightarrow d$.
The gluon is taken to have a four-momentum given by $p_e$, and helicity
$\lambda_e$.  The momenta of the $W^+$ and $W^-$ final-state
bosons are denoted by $q_1$ and $q_2$, respectively.


The $q\bar{q}q\bar{q}W^+W^-g$ amplitudes are evaluated numerically using the
HELAS program package~\cite{HELAS}. The HELAS program contains subroutines
evaluating and multiplying the various factors in the Feynman graphs. Hence we
shall employ a notation directly relating to the structure of HELAS.
To begin, the generalized propagator for
a vector boson of mass $M$ in the unitary gauge is given by
\begin{equation}
D^{\mu \nu}(q^2, M) = {-i \over q^2 - M^2 + i M \Gamma\, \Theta(q^2)}
   \left[ g^{\mu \nu} - {q^\mu q^\nu \over M^2} \right]
\end{equation}
with $q^2$ representing the momentum transfer squared, and $\Gamma$ denoting
the $q^2$ independent decay width of the intermediate particle. The step
function $\Theta(q^2)$ eliminates the imaginary part of the vector boson
propagator for spacelike momentum transfer.
The analogous choice for the heavy Higgs boson propagator will introduce only
a small error compared to unitarity requirements~\cite{Will90}.
Similarly, for massless particles, the propagator function is defined
as
\begin{equation}
d^{\mu \nu}(q^2) = -{i \over q^2} g^{\mu \nu}
\end{equation}

Vertex insertions are defined via
\begin{equation}
V_i^\mu = \gamma^\mu \left[ g_V^i + g_A^i \gamma^5 \right]
\end{equation}
where the index $i$ on the vector and axial coupling constants $g_V^i$ and
$g_A^i$ will list the vector bosons and fermions coupling at a particular
vertex.

The zero-width-approximation is employed in order to include
spin-correlation effects of the $W^\pm$ decay products. The polarization
vectors of the final state $W$ bosons are denoted by $\epsilon^\ast_1(q_1)$
for the $W^+$ boson and $\epsilon^\ast_2(q_2)$ for the $W^-$ boson.
\begin{eqnarray}
\epsilon^\mu_1(q_1, \lambda_1)^\ast &=& J^\mu_{12}(k_1, k_2, \rho_1, \rho_2)
   \sqrt{\pi \over {M_W \Gamma_W}} \\
\epsilon^\mu_2(q_2, \lambda_2)^\ast &=& J^\mu_{34}(k_3, k_4, \rho_3, \rho_4)
   \sqrt{\pi \over {M_W \Gamma_W}}
\end{eqnarray}
where the truncated decay current is defined in terms of the final state
fermion spinors
\begin{eqnarray}
J^\mu_{12} &=& \bar{u}(k_1, \rho_1) V^\mu_{Wff} v(k_2, \rho_2) \\
J^\mu_{34} &=& \bar{u}(k_3, \rho_3) V^\mu_{Wff} v(k_4, \rho_4)
\end{eqnarray}
Inserting these decay currents for the $W$ polarization vectors the resulting
matrix elements will describe the full process
\begin{equation}\label{procapp}
q_aq_c\to q_bq_dW^+W^-g\; ,\qquad  W^+\to \ell_1\bar{\ell_2}\;,\quad
W^-\to \ell_3\bar{\ell_4}\;,
\end{equation}
and all processes related by crossing of the quarks and gluons.

In order to simplify the spinor algebra we use a bra and ket notation:
\begin{eqnarray}
u(p_a,\lambda_a) &=& |a\bigr> \\
u(p_b,\lambda_b) &=& |b\bigr> \\
\bar{u}(p_c,\lambda_c) &=& \bigl<c| \\
\bar{u}(p_d,\lambda_d) &=& \bigl<d|
\end{eqnarray}
and similarly, the wavefunctions for antiquarks are defined as
\begin{eqnarray}
\bar{v}(p_a,-\lambda_a) &=& \bigl<a| \\
\bar{v}(p_b,-\lambda_b) &=& \bigl<b| \\
v(p_c,-\lambda_c) &=& |c\bigr> \\
v(p_d,-\lambda_d) &=& |d\bigr>
\end{eqnarray}

The currents, corresponding to the splitting of a virtual $\gamma,Z$ into
a final-state $W^+W^-$ pair via the triple-boson vertex, are
\begin{eqnarray}
U^\mu_\gamma &=&
   d^{\mu \nu}((q_1+q_2)^2)
   E_{\rho \lambda \nu}
   \epsilon^{\lambda \ast}_1 \epsilon^{\rho \ast}_2 \\
U^\mu_Z &=&
   D^{\mu \nu}((q_1+q_2)^2,M_Z)
   F_{\rho \lambda \nu}
   \epsilon^{\lambda \ast}_1 \epsilon^{\rho \ast}_2
\end{eqnarray}
where $E$ represents the $WW\gamma$ triple-boson vertex rule, with the index
ordering corresponding to incoming $W^+W^-\gamma$ bosons or outgoing
$W^-W^+\gamma$ bosons.  The tensor $F$ represents the $WWZ$ triple-boson vertex
rule in analogous notation.

Off-shell fermion wavefunctions, corresponding to gauge boson emission off the
fermion lines,  are defined via
\begin{eqnarray}
|1,a\bigr> &=&
   {i \over \sla p_a - \sla q_1} \epsilon^{\mu\ast}_1 V^{Wff}_\mu |a\bigr> \\
|2,1,a\bigr> &=&
   {i \over \sla p_a - \sla q_1 - \sla q_2} \epsilon^{\mu\ast}_2
   V^{Wff}_\mu |1,a\bigr> \\
|U,a\bigr> &=&
   {i \over \sla p_a - \sla q_1 - \sla q_2}
   \left( U^\mu_\gamma V^{\gamma ff}_\mu + U^\mu_Z V^{Zff}_\mu \right)
   |a\bigr> \\
\bigl<c,2| &=&
   \bigl<c| V^{Wff}_\mu \epsilon^{\mu\ast}_2 {i \over \sla p_c + \sla q_2} \\
\bigl<c,2,1| &=&
   \bigl<c,2| V^{Wff}_\mu \epsilon^{\mu\ast}_1 {i \over \sla p_c + \sla q_2 +
\sla q_1} \\
\bigl<c,U| &=&
   \bigl<c|
   \left( U^\mu_\gamma V^{\gamma ff}_\mu + U^\mu_Z V^{Zff}_\mu \right)
   {i \over \sla p_c + \sla q_2 + \sla q_1}
\end{eqnarray}
Similar expressions hold for wavefunctions involving fermions $b$ and $d$.

Gluon insertions along a fermion line are notated as
\begin{eqnarray}
|g,a\bigr> &=&
   {i \over \sla p_a - \sla p_e} \epsilon^{\mu\ast}_g V^{gff}_\mu |a\bigr> \\
|1,g,a\bigr> &=&
   {i \over \sla p_a - \sla p_e - \sla q_1} \epsilon^{\mu\ast}_1
   V^{Wff}_\mu |g,a\bigr> \\
|g,1,a\bigr> &=&
   {i \over \sla p_a - \sla p_e - \sla q_1} \epsilon^{\mu\ast}_g
   V^{gff}_\mu |1,a\bigr> \\
|2,g,1,a\bigr> &=&
   {i \over \sla p_a - \sla p_e - \sla q_1 - \sla q_2} \epsilon^{\mu\ast}_2
   V^{Wff}_\mu |g,1,a\bigr> \\
|g,2,1,a\bigr> &=&
   {i \over \sla p_a - \sla p_e - \sla q_1 - \sla q_2} \epsilon^{\mu\ast}_g
   V^{gff}_\mu |2,1,a\bigr> \\
\bigl<c,g| &=&
   \bigl<c| V^{gff}_\mu \epsilon^{\mu\ast}_g {i \over \sla p_c + \sla p_e} \\
\bigl<c,g,2| &=&
   \bigl<c,g| V^{Wff}_\mu \epsilon^{\mu\ast}_2 {i \over \sla p_c +
\sla p_e + \sla q_2} \\
\bigl<c,2,g| &=&
   \bigl<c,2| V^{gff}_\mu \epsilon^{\mu\ast}_g {i \over \sla p_c +
\sla p_e + \sla q_2} \\
\bigl<c,2,g,1| &=&
   \bigl<c,2,g| V^{Wff}_\mu \epsilon^{\mu\ast}_1 {i \over \sla p_c +
\sla p_e + \sla q_1 + \sla q_2} \\
\bigl<c,2,1,g| &=&
   \bigl<c,2,1| V^{gff}_\mu \epsilon^{\mu\ast}_g {i \over \sla p_c +
\sla p_e + \sla q_1 + \sla q_2}
\end{eqnarray}

Simple electroweak currents are denoted as in
\begin{eqnarray}
J^\mu_{\gamma}(ac) &=&
   d^{\mu \nu}((p_a - p_c)^2) \bigl<c| V^{\gamma ff}_\nu |a\bigr> \\
J^\mu_{W}(ac) &=&
   D^{\mu \nu}((p_a - p_c)^2,M_W) \bigl<c| V^{Wff}_\nu |a\bigr> \\
J^\mu_{Z}(ac) &=&
   D^{\mu \nu}((p_a - p_c)^2,M_Z) \bigl<c| V^{Zff}_\nu |a\bigr> \\
J^\mu_{3}(ac) &=&
   \cos\theta_W J^\mu_Z(ac) + \sin\theta_W J^\mu_\gamma(ac)
\end{eqnarray}
with $(ac)$ representing fermion current flow from $a \rightarrow c$. Fermion
current flow from $c \rightarrow a$ would be represented by the argument
$(ca)$.

When including gluon emission the notation
\begin{eqnarray}
J^\mu_{g\gamma}(ac) &=&
   d^{\mu \nu}((p_a - p_c - p_e)^2) \bigl<c| V^{\gamma ff}_\nu |g,a\bigr> \\
J^\mu_{\gamma g}(ac) &=&
   d^{\mu \nu}((p_a - p_c - p_e)^2) \bigl<c,g| V^{\gamma ff}_\nu |a\bigr> \\
J^\mu_{g\gamma}(ca) &=&
   d^{\mu \nu}((p_c - p_a - p_e)^2) \bigl<a,g| V^{\gamma ff}_\nu |c\bigr> \\
J^\mu_{\gamma g}(ca) &=&
   d^{\mu \nu}((p_c - p_a - p_e)^2) \bigl<a| V^{\gamma ff}_\nu |g,c\bigr>
\end{eqnarray}
is used, and similarly more complicated single $W$ boson emission currents are
denoted by
\begin{eqnarray}
J^\mu_{1\gamma}(ac) &=&
   d^{\mu \nu}((p_a - p_c - q_1)^2)
   \bigl<c| V^{\gamma ff}_\nu |1,a\bigr> \\
J^\mu_{1W}(ac) &=&
   D^{\mu \nu}((p_a - p_c - q_1)^2,M_W)
   \bigl<c| V^{Wff}_\nu |1,a\bigr> \\
J^\mu_{Z1}(ac) &=&
   D^{\mu \nu}((p_a - p_c - q_1)^2,M_Z)
   \bigl<c,1| V^{Zff}_\nu |a\bigr>\; .
\end{eqnarray}
Including the emission of a gluon, the notation of the latter three changes to
\begin{eqnarray}
J^\mu_{g\gamma 1}(ac) &=&
   d^{\mu \nu}((p_a - p_c - p_e - q_1)^2)
   \bigl<c,1| V^{\gamma ff}_\nu |g,a\bigr> \\
J^\mu_{Wg1}(ac) &=&
   D^{\mu \nu}((p_a - p_c - p_e - q_1)^2,M_W)
   \bigl<c,1,g| V^{Wff}_\nu |a\bigr> \\
J^\mu_{Z1g}(ac) &=&
   D^{\mu \nu}((p_a - p_c - p_e - q_1)^2,M_Z)
   \bigl<c,g,1| V^{Zff}_\nu |a\bigr>
\end{eqnarray}
and similarly for interchange of outgoing bosons 1 and 2, interchange
of fermion current flow $a \leftrightarrow c$, interchange of fermion
line $(ac)$ for $(bd)$, or a combination of any and/or all of the above
interchanges. Notice that in all cases the space-time index of the current
corresponds to the {\it explicitly} listed {\it electroweak} boson in the
index identifying the current.

Double $W$ boson emission currents may be defined similarly, as in the examples
\begin{eqnarray}
J^\mu_{12\gamma}(ac) &=&
   d^{\mu \nu}((p_a - p_c - q_1 - q_2)^2)
   \bigl<c| V^{\gamma ff}_\nu |2,1,a\bigr> \\
J^\mu_{1W2}(ac) &=&
   D^{\mu \nu}((p_a - p_c - q_1 - q_2)^2,M_W)
   \bigl<c,2| V^{Wff}_\nu |1,a\bigr> \\
J^\mu_{ZU}(ac) &=&
   D^{\mu \nu}((p_a - p_c - q_1 - q_2)^2,M_Z)
   \bigl<c,U| V^{Zff}_\nu |a\bigr>\; .
\end{eqnarray}
Including gluon emission another index needs to be added,
\begin{eqnarray}
J^\mu_{1g\gamma 2}(ac) &=&
   d^{\mu \nu}((p_a - p_c - p_e - q_1 - q_2)^2)
   \bigl<c,2| V^{\gamma ff}_\nu |g,1,a\bigr> \\
J^\mu_{WgU}(ac) &=&
   D^{\mu \nu}((p_a - p_c - p_e - q_1 - q_2)^2,M_W)
   \bigl<c,U,g| V^{Wff}_\nu |a\bigr> \\
J^\mu_{gZ12}(ac) &=&
   D^{\mu \nu}((p_a - p_c - p_e - q_1 - q_2)^2,M_Z)
   \bigl<c,2,1| V^{Zff}_\nu |g,a\bigr>\; ,
\end{eqnarray}
where again, all possible interchanges of outgoing bosons, fermion
current flows, and fermion lines are allowed.

For sums of currents involving gluon emission off the same fermion
line, a semi-colon notation is introduced.  For example,
\begin{eqnarray}
J^\mu_{\gamma;g}(ac) &=& J^\mu_{g\gamma}(ac) + J^\mu_{\gamma g}(ac) \\
J^\mu_{Z12;g}(ac) &=& J^\mu_{gZ12}(ac) + J^\mu_{Zg12}(ac)
   + J^\mu_{Z1g2}(ac) + J^\mu_{Z12g}(ac)
\end{eqnarray}
represent a photon and a $Z$ current given by the sum of currents involving
a gluon emission along {\it all} possible topologically distinct locations
on the fermion line.  In terms of the bra and ket notation, the semi-colon
notation is analogously introduced by example,
\begin{eqnarray}
\bigl<c,1,2| X |a\bigr>_{;g}
   &=& \bigl<c,1,2| X |g,a\bigr> + \bigl<c,1,2,g| X |a\bigr> \nonumber \\
   &+& \bigl<c,1,g,2| X |a\bigr> + \bigl<c,g,1,2| X |a\bigr>\; ,
\end{eqnarray}
where $X$ is an arbitrary vertex insertion.

The scalar current obtained from $WW$ fusion into a Higgs boson
is defined via
\begin{equation}
H^{WWH}_{\mu \nu} =
   D((q_1+q_2)^2,M_H) I^{WWH}_{\mu \nu}
\end{equation}
where $I^{WWH}_{\mu \nu}$ represents the $WWH$ vertex rule.  Similarly,
$I^{ZZH}_{\mu \nu}$ represents the $ZZH$ vertex rule.

The $SU(2)$ quadruple-boson vertices are represented by $S$ and $T$.
$S_{\mu \nu \lambda \rho}$ represents the $WWWW$ effective coupling,
which includes the contact term, as well as $s$-channel and $t$-channel
$\gamma,Z$ exchange terms.
$T_{\mu \nu \lambda \rho}$ represents the $WW^3WW^3$ effective coupling,
which includes the contact term, as well as $t$-channel and $u$-channel
$W^\pm$ exchange terms.

The tensor $G$ represents the $WWW^3$ triple-boson vertex rule, with
the index ordering corresponding to incoming $W^+W^-W^3$ bosons, or outgoing
$W^-W^+W^3$ bosons, respectively.  The current $J_3$ is, as defined in
Eq.~(A38), a $SU(2)$ superposition of $\gamma,Z$ currents.

With this notation we can now express the amplitudes, which correspond to the
various Feynman graphs, in a very compact notation which can directly be
translated into a calling sequence for HELAS subroutines.
For the Feynman graphs of Figure~\ref{FigFeynAB}.a one gets
\begin{eqnarray}
-i{\cal M}_{1}(a) &=&
   J^\mu_{W;g}(ac) J^\nu_W(bd)
   I_{\mu \nu}^{WWH} H_{\lambda \rho}^{WWH}
   \epsilon^{\lambda\ast}_1 \epsilon^{\rho\ast}_2 \\
-i{\cal M}_{1}(b) &=&
   J^\nu_{W;g}(bd) J^\mu_W(ac)
   I_{\mu \nu}^{WWH} H_{\lambda \rho}^{WWH}
   \epsilon^{\lambda\ast}_1 \epsilon^{\rho\ast}_2
\end{eqnarray}
where ${\cal M}(a)$ represents an amplitude including all distinguishable
gluon emission insertions along the fermion line $a \rightarrow c$.
Similarly, ${\cal M}(b)$ represents an amplitude including all
distinguishable gluon emission insertions along the fermion line
$b \rightarrow d$.

Only the ${\cal M}(a)$ amplitudes are shown below,
with the ${\cal M}(b)$ amplitudes derived via interchange of the
fermion indices ($ac \rightarrow bd$). The remaining ${\cal M}(a)$ amplitudes
can be represented as follows
\begin{eqnarray}
-i{\cal M}_{2}(a) &=&
   J^\mu_{W;g}(ac) J^\nu_W(bd)
   I_{\mu \lambda}^{WWH} H_{\nu \rho}^{WWH}
   \epsilon^{\lambda\ast}_1 \epsilon^{\rho\ast}_2 \\
-i{\cal M}_{3}(a) &=&
   J^\mu_{W;g}(ac) J^\nu_W(bd)
   I_{\mu \rho}^{WWH} H_{\nu \lambda}^{WWH}
   \epsilon^{\lambda\ast}_1 \epsilon^{\rho\ast}_2
\end{eqnarray}
with corresponding Feynman diagrams presented in Figure~\ref{FigFeynAB}.b.
Notice that for a given set of external quark flavors only one of the two
amplitudes ${\cal M}_{2}$ or ${\cal M}_{3}$ will be nonvanishing:
${\cal M}_{2}$ corresponds to a $W^+$ coupling to the $(ac)$ quark line
while for ${\cal M}_{3}$ the emission of a $W^-$ off the $(ac)$ quark line must
be allowed. Analogous flavor selection rules will be implied for all remaining
diagrams as well. The amplitudes for these are given by
\begin{eqnarray}
-i{\cal M}_{4}(a) &=&
   J^\mu_{W;g}(ac) J^\nu_W(bd)
   S_{\mu \nu \rho \lambda}
   \epsilon^{\lambda\ast}_1 \epsilon^{\rho\ast}_2 \\
-i{\cal M}_{5}(a) &=&
   J^\mu_{W;g}(ac) J^\nu_W(bd)
   S_{\nu \mu \rho \lambda}
   \epsilon^{\lambda\ast}_1 \epsilon^{\rho\ast}_2 \\
-i{\cal M}_{6}(a) &=&
   J^\mu_{Z;g}(ac) J^\nu_Z(bd)
   I_{\mu \nu}^{ZZH} H_{\lambda \rho}^{WWH}
   \epsilon^{\lambda\ast}_1 \epsilon^{\rho\ast}_2 \\
-i{\cal M}_{7}(a) &=&
   J^\mu_{3;g}(ac) J^\nu_3(bd)
   T_{\rho \mu \lambda \nu}
   \epsilon^{\lambda\ast}_1 \epsilon^{\rho\ast}_2 \\
\noalign{\hbox{\rm with corresponding Feynman diagrams presented in
Figure~\ref{FigFeynCE} }}
-i{\cal M}_{8}(a) &=&
   J^\mu_{2W;g}(ac) J^\nu_3(bd)
   G_{\mu \lambda \nu}
   \epsilon^{\lambda \ast}_1 \\
-i{\cal M}_{9}(a) &=&
   J^\mu_{W2;g}(ac) J^\nu_3(bd)
   G_{\mu \lambda \nu}
   \epsilon^{\lambda \ast}_1 \\
-i{\cal M}_{10}(a) &=&
   J^\mu_{1W;g}(ac) J^\nu_3(bd)
   G_{\lambda \mu \nu}
   \epsilon^{\lambda \ast}_2 \\
-i{\cal M}_{11}(a) &=&
   J^\mu_{W1;g}(ac) J^\nu_3(bd)
   G_{\lambda \mu \nu}
   \epsilon^{\lambda \ast}_2 \\
-i{\cal M}_{12}(a) &=&
   J^\mu_{3;g}(ac) J^\nu_{2W}(bd)
   G_{\nu \lambda \mu}
   \epsilon^{\lambda \ast}_1 \\
-i{\cal M}_{13}(a) &=&
   J^\mu_{3;g}(ac) J^\nu_{W2}(bd)
   G_{\nu \lambda \mu}
   \epsilon^{\lambda \ast}_1 \\
-i{\cal M}_{14}(a) &=&
   J^\mu_{3;g}(ac) J^\nu_{1W}(bd)
   G_{\lambda \nu \mu}
   \epsilon^{\lambda \ast}_2 \\
-i{\cal M}_{15}(a) &=&
   J^\mu_{3;g}(ac) J^\nu_{W1}(bd)
   G_{\lambda \nu \mu}
   \epsilon^{\lambda \ast}_2 \\
-i{\cal M}_{16}(a) &=&
   J^\mu_{W;g}(ac) J^\nu_{23}(bd)
   G_{\mu \lambda \nu}
   \epsilon^{\lambda \ast}_1 \\
-i{\cal M}_{17}(a) &=&
   J^\mu_{W;g}(ac) J^\nu_{32}(bd)
   G_{\mu \lambda \nu}
   \epsilon^{\lambda \ast}_1 \\
-i{\cal M}_{18}(a) &=&
   J^\mu_{W;g}(ac) J^\nu_{13}(bd)
   G_{\lambda \mu \nu}
   \epsilon^{\lambda \ast}_2 \\
-i{\cal M}_{19}(a) &=&
   J^\mu_{W;g}(ac) J^\nu_{31}(bd)
   G_{\lambda \mu \nu}
   \epsilon^{\lambda \ast}_2 \\
-i{\cal M}_{20}(a) &=&
   J^\mu_{23;g}(ac) J^\nu_{W}(bd)
   G_{\nu \lambda \mu}
   \epsilon^{\lambda \ast}_1 \\
-i{\cal M}_{21}(a) &=&
   J^\mu_{32;g}(ac) J^\nu_{W}(bd)
   G_{\nu \lambda \mu}
   \epsilon^{\lambda \ast}_1 \\
-i{\cal M}_{22}(a) &=&
   J^\mu_{13;g}(ac) J^\nu_{W}(bd)
   G_{\lambda \nu \mu}
   \epsilon^{\lambda \ast}_2 \\
-i{\cal M}_{23}(a) &=&
   J^\mu_{31;g}(ac) J^\nu_{W}(bd)
   G_{\lambda \nu \mu}
   \epsilon^{\lambda \ast}_2 \\
-i{\cal M}_{24}(a) &=&
   J^\mu_{12\gamma;g}(ac) \bigl<d| V^{\gamma ff}_\mu |b\bigr>
 + J^\mu_{12Z;g}(ac) \bigl<d| V^{Zff}_\mu |b\bigr> \\
-i{\cal M}_{25}(a) &=&
   J^\mu_{1\gamma2;g}(ac) \bigl<d| V^{\gamma ff}_\mu |b\bigr>
 + J^\mu_{1Z2;g}(ac) \bigl<d| V^{Zff}_\mu |b\bigr> \\
-i{\cal M}_{26}(a) &=&
   J^\mu_{\gamma12;g}(ac) \bigl<d| V^{\gamma ff}_\mu |b\bigr>
 + J^\mu_{Z12;g}(ac) \bigl<d| V^{Zff}_\mu |b\bigr> \\
-i{\cal M}_{27}(a) &=&
   J^\mu_{21\gamma;g}(ac) \bigl<d| V^{\gamma ff}_\mu |b\bigr>
 + J^\mu_{21Z;g}(ac) \bigl<d| V^{Zff}_\mu |b\bigr> \\
-i{\cal M}_{28}(a) &=&
   J^\mu_{2\gamma1;g}(ac) \bigl<d| V^{\gamma ff}_\mu |b\bigr>
 + J^\mu_{2Z1;g}(ac) \bigl<d| V^{Zff}_\mu |b\bigr> \\
-i{\cal M}_{29}(a) &=&
   J^\mu_{\gamma21;g}(ac) \bigl<d| V^{\gamma ff}_\mu |b\bigr>
 + J^\mu_{Z21;g}(ac) \bigl<d| V^{Zff}_\mu |b\bigr> \\
-i{\cal M}_{30}(a) &=&
   J^\mu_{12\gamma}(bd) \bigl<c| V^{\gamma ff}_\mu |a\bigr>_{;g}
 + J^\mu_{12Z}(bd) \bigl<c| V^{Zff}_\mu |a\bigr>_{;g} \\
-i{\cal M}_{31}(a) &=&
   J^\mu_{1\gamma2}(bd) \bigl<c| V^{\gamma ff}_\mu |a\bigr>_{;g}
 + J^\mu_{1Z2}(bd) \bigl<c| V^{Zff}_\mu |a\bigr>_{;g} \\
-i{\cal M}_{32}(a) &=&
   J^\mu_{\gamma12}(bd) \bigl<c| V^{\gamma ff}_\mu |a\bigr>_{;g}
 + J^\mu_{Z12}(bd) \bigl<c| V^{Zff}_\mu |a\bigr>_{;g} \\
-i{\cal M}_{33}(a) &=&
   J^\mu_{21\gamma}(bd) \bigl<c| V^{\gamma ff}_\mu |a\bigr>_{;g}
 + J^\mu_{21Z}(bd) \bigl<c| V^{Zff}_\mu |a\bigr>_{;g} \\
-i{\cal M}_{34}(a) &=&
   J^\mu_{2\gamma1}(bd) \bigl<c| V^{\gamma ff}_\mu |a\bigr>_{;g}
 + J^\mu_{2Z1}(bd) \bigl<c| V^{Zff}_\mu |a\bigr>_{;g} \\
-i{\cal M}_{35}(a) &=&
   J^\mu_{\gamma21}(bd) \bigl<c| V^{\gamma ff}_\mu |a\bigr>_{;g}
 + J^\mu_{Z21}(bd) \bigl<c| V^{Zff}_\mu |a\bigr>_{;g} \\
\noalign{\hbox{\rm with corresponding Feynman diagrams presented in
Figure~\ref{FigFeynFH} }}
-i{\cal M}_{36}(a) &=&
   J^\mu_{1\gamma;g}(ac) \bigl<d| V^{\gamma ff}_\mu |2,b\bigr>
 + J^\mu_{1Z;g}(ac) \bigl<d| V^{Zff}_\mu |2,b\bigr> \\
-i{\cal M}_{37}(a) &=&
   J^\mu_{\gamma1;g}(ac) \bigl<d| V^{\gamma ff}_\mu |2,b\bigr>
 + J^\mu_{Z1;g}(ac) \bigl<d| V^{Zff}_\mu |2,b\bigr> \\
-i{\cal M}_{38}(a) &=&
   J^\mu_{1\gamma;g}(ac) \bigl<d,2| V^{\gamma ff}_\mu |b\bigr>
 + J^\mu_{1Z;g}(ac) \bigl<d,2| V^{Zff}_\mu |b\bigr> \\
-i{\cal M}_{39}(a) &=&
   J^\mu_{\gamma1;g}(ac) \bigl<d,2| V^{\gamma ff}_\mu |b\bigr>
 + J^\mu_{Z1;g}(ac) \bigl<d,2| V^{Zff}_\mu |b\bigr> \\
-i{\cal M}_{40}(a) &=&
   J^\mu_{2\gamma;g}(ac) \bigl<d| V^{\gamma ff}_\mu |1,b\bigr>
 + J^\mu_{2Z;g}(ac) \bigl<d| V^{Zff}_\mu |1,b\bigr> \\
-i{\cal M}_{41}(a) &=&
   J^\mu_{\gamma2;g}(ac) \bigl<d| V^{\gamma ff}_\mu |1,b\bigr>
 + J^\mu_{Z2;g}(ac) \bigl<d| V^{Zff}_\mu |1,b\bigr> \\
-i{\cal M}_{42}(a) &=&
   J^\mu_{2\gamma;g}(ac) \bigl<d,1| V^{\gamma ff}_\mu |b\bigr>
 + J^\mu_{2Z;g}(ac) \bigl<d,1| V^{Zff}_\mu |b\bigr> \\
-i{\cal M}_{43}(a) &=&
   J^\mu_{\gamma2;g}(ac) \bigl<d,1| V^{\gamma ff}_\mu |b\bigr>
 + J^\mu_{Z2;g}(ac) \bigl<d,1| V^{Zff}_\mu |b\bigr> \\
-i{\cal M}_{44}(a) &=&
   J^\mu_{U\gamma;g}(ac) \bigl<d| V^{\gamma ff} |b\bigr>
 + J^\mu_{UZ;g}(ac) \bigl<d| V^{Zff} |b\bigr> \\
-i{\cal M}_{45}(a) &=&
   J^\mu_{\gamma U;g}(ac) \bigl<d| V^{\gamma ff} |b\bigr>
 + J^\mu_{ZU;g}(ac) \bigl<d| V^{Zff} |b\bigr> \\
-i{\cal M}_{46}(a) &=&
   J^\mu_{U\gamma}(bd) \bigl<c| V^{\gamma ff} |a\bigr>_{;g}
 + J^\mu_{UZ}(bd) \bigl<c| V^{Zff} |a\bigr>_{;g} \\
-i{\cal M}_{47}(a) &=&
   J^\mu_{\gamma U}(bd) \bigl<c| V^{\gamma ff} |a\bigr>_{;g}
 + J^\mu_{ZU}(bd) \bigl<c| V^{Zff} |a\bigr>_{;g} \\
-i{\cal M}_{48}(a) &=&
   J^\mu_{12W;g}(ac) \bigl<d| V^{Wff}_\mu |b\bigr> \\
-i{\cal M}_{49}(a) &=&
   J^\mu_{21W;g}(ac) \bigl<d| V^{Wff}_\mu |b\bigr> \\
-i{\cal M}_{50}(a) &=&
   J^\mu_{W21;g}(ac) \bigl<d| V^{Wff}_\mu |b\bigr> \\
-i{\cal M}_{51}(a) &=&
   J^\mu_{W12;g}(ac) \bigl<d| V^{Wff}_\mu |b\bigr> \\
-i{\cal M}_{52}(a) &=&
   J^\mu_{21W}(bd) \bigl<c| V^{Wff}_\mu |a\bigr>_{;g} \\
-i{\cal M}_{53}(a) &=&
   J^\mu_{12W}(bd) \bigl<c| V^{Wff}_\mu |a\bigr>_{;g} \\
-i{\cal M}_{54}(a) &=&
   J^\mu_{W12}(bd) \bigl<c| V^{Wff}_\mu |a\bigr>_{;g} \\
-i{\cal M}_{55}(a) &=&
   J^\mu_{W21}(bd) \bigl<c| V^{Wff}_\mu |a\bigr>_{;g} \\
\noalign{\hbox{\rm with corresponding Feynman diagrams presented in
Figure~\ref{FigFeynIK}, and finally }}
-i{\cal M}_{56}(a) &=&
   J^\mu_{2W;g}(ac) \bigl<d| V^{Wff}_\mu |1,b\bigr> \\
-i{\cal M}_{57}(a) &=&
   J^\mu_{W2;g}(ac) \bigl<d| V^{Wff}_\mu |1,b\bigr> \\
-i{\cal M}_{58}(a) &=&
   J^\mu_{2W;g}(ac) \bigl<d,1| V^{Wff}_\mu |b\bigr> \\
-i{\cal M}_{59}(a) &=&
   J^\mu_{W2;g}(ac) \bigl<d,1| V^{Wff}_\mu |b\bigr> \\
-i{\cal M}_{60}(a) &=&
   J^\mu_{1W;g}(ac) \bigl<d| V^{Wff}_\mu |2,b\bigr> \\
-i{\cal M}_{61}(a) &=&
   J^\mu_{W1;g}(ac) \bigl<d| V^{Wff}_\mu |2,b\bigr> \\
-i{\cal M}_{62}(a) &=&
   J^\mu_{1W;g}(ac) \bigl<d,2| V^{Wff}_\mu |b\bigr> \\
-i{\cal M}_{63}(a) &=&
   J^\mu_{W1;g}(ac) \bigl<d,2| V^{Wff}_\mu |b\bigr> \\
-i{\cal M}_{64}(a) &=&
   J^\mu_{UW;g}(ac) \bigl<d| V^{Wff} |b\bigr> \\
-i{\cal M}_{65}(a) &=&
   J^\mu_{WU;g}(ac) \bigl<d| V^{Wff} |b\bigr> \\
-i{\cal M}_{66}(a) &=&
   J^\mu_{UW}(bd) \bigl<c| V^{Wff} |a\bigr>_{;g} \\
-i{\cal M}_{67}(a) &=&
   J^\mu_{WU}(bd) \bigl<c| V^{Wff} |a\bigr>_{;g}
\end{eqnarray}
with corresponding Feynman diagrams presented in Figure~\ref{FigFeynLM}.
Notice that there are no diagrams involving Goldstone bosons.  This is a
remnant of the choice of the unitary gauge for this calculation.

Once the individual amplitudes are evaluated, the two amplitude sums
are calculated as
\begin{eqnarray}
   {\cal M}(a) &=& {\cal M}_1(a) + {\cal M}_2(a) + \ldots + {\cal M}_{67}(a) \\
   {\cal M}(b) &=& {\cal M}_1(b) + {\cal M}_2(b) + \ldots + {\cal M}_{67}(b)
\end{eqnarray}
The overall amplitude including the QCD color structure is hence
\begin{equation}
   {\cal M} = T^i_{ac} \delta_{bd} {\cal M}(a)
            + \delta_{ac} T^i_{bd} {\cal M}(b)
\label{EqnCSas}
\end{equation}
where $a, b, c$ and $d$ represent the color indices of the four
quarks, and $i$ represents the color index of the attached gluon.
The required sum of the amplitude square, $|{\cal M}|^2$, over all
external colors thus yields
\begin{equation}\label{colorsum}
\sum_{\rm colors}|{\cal M}|^2 = N{N^2-1\over 2} \;
\left(\; |{\cal M}(a)|^2 +  |{\cal M}(b)|^2 \; \right)
\end{equation}
for an $SU(N)$ gauge theory ($N=3$ for QCD).

When identical final-state fermions are present one must anti-symmetrize
the resulting matrix element.  This complicates the resulting color
structure.  In comparison with Eqn.~\ref{EqnCSas} above, the resulting
amplitude contains both $t$ and $u$-channel processes as
\begin{eqnarray}
   {\cal M} &=& T^i_{ac} \delta_{bd} {\cal M}(a)
            + \delta_{ac} T^i_{bd} {\cal M}(b) \nonumber \\
   &-& T^i_{ad} \delta_{bc} {\cal M}(a)[c\leftrightarrow d]
            - \delta_{ad} T^i_{bc} {\cal M}(b)[c\leftrightarrow d]
\label{EqnCSasp}
\end{eqnarray}
This leads to  modified terms in the analogue of Eq.~(\ref{colorsum}). Defining
\begin{eqnarray}
   {\cal M}_1' &=& {\cal M}(a) + {\cal M}(b)
      - {\cal M}(a)[c \leftrightarrow d] - {\cal M}(b)[c \leftrightarrow d] \\
   {\cal M}_2' &=& {\cal M}(a) - {\cal M}(b) \\
   {\cal M}_3' &=&
      {\cal M}(a)[c \leftrightarrow d] - {\cal M}(b)[c \leftrightarrow d] \\
   {\cal M}_4' &=& {\cal M}(a) + {\cal M}(b)
      + {\cal M}(a)[c \leftrightarrow d] + {\cal M}(b)[c \leftrightarrow d]
\end{eqnarray}
one obtains
\begin{equation}\label{colorsumtu}
\sum_{\rm colors}|{\cal M}|^2 = \sum_{j=1}^4\; C_j'\; |{\cal M}_j'|^2
\end{equation}
with
\begin{eqnarray}
   C_1' &=& (N+2) \left( {N^2 - 1 \over 2} \right) \\
   C_2' &=& N \left( {N^2 - 1 \over 2} \right) \\
   C_3' &=& N \left( {N^2 - 1 \over 2} \right) \\
   C_4' &=& (N-2) \left( {N^2 - 1 \over 2} \right)\; .
\end{eqnarray}
Within $SU(3)$ one needs to insert the color factors $C_1' = 20$,
$C_2' = C_3' = 12$ and $C_4' = 4$.
Inclusion of $s$-channel diagrams, and possible $t$ and $s$-channel
interference proceeds with identical color structures as for the $t$-channel,
and $t$ and $u$-channel interference cases, respectively.

\begin{table}
\caption{Values (in GeV) of the $p_{tsa}$ parameter of Eq.~(\protect\ref{reg})
which are needed to match the cross sections at ${\cal O}(\alpha_s)$ with the
lowest order results for the ``$s$-channel resonance'' process. Values are
given for two Higgs boson masses and different transverse momentum requirements
for ``visible'' jets in $pp$ collisions at $\protect\sqrt{s}=40$~TeV.}
\begin{tabular}{lcccc}
& \multicolumn{2}{c}{1 jet inclusive}& \multicolumn{2}{c}{2 jet inclusive}\\
& $m_H=500$ GeV& $m_H=800$ GeV& $m_H=500$ GeV& $m_H=800$ GeV\\
$p_{Tj} > 20$ GeV& 5.8& 5.7& 6.7& 6.5\\
$p_{Tj} > 40$ GeV& 6.1& 5.9& 7.9& 7.6\\
$p_{Tj} > 60$ GeV& 6.4& 6.3& 8.9& 8.9
\end{tabular}
\end{table}

\begin{table}
\caption{Cross sections (in fb) and central-jet-veto efficiencies $\varepsilon$
for $pp$ collisions at $\protect\sqrt s=40$~TeV. Only leptonic decays of both
$W$'s ($W^\pm\to e^\pm\nu, \mu^\pm\nu$) are considered and the lepton and
tagging jet cuts of Eqs.~(\protect\ref{cutell2})--(\protect\ref{Ejtag}) are
imposed, leaving $B\cdot \sigma (m_H=800{\rm~GeV}) = 8.2(7.6)$~fb and
$B\cdot \sigma (m_H=100{\rm~GeV}) = 1.9(1.5)$~fb before central-jet-vetoing
for the ${\cal O}(\alpha_s)$ (lowest
order) calculation. A veto is imposed on central jets of $|\eta_j|<3$ and
transverse momenta $p_{Tj}>p_T^{\rm veto}$ as given in the table. The signal is
defined as the difference of the $m_H=800$~GeV and $m_H=100$~GeV cross
sections.}
\begin{tabular}{rcccc}
\multicolumn{5}{c}{a) 2 parton calculation}\\
& $B\cdot\sigma(m_H=800~\rm GeV)$& $B\cdot\sigma(m_H=100\rm~GeV)$&
$B\cdot\sigma_{\rm signal}$& $\varepsilon$\\
$p_T^{\rm veto}=20$ GeV& 4.4& 0.5& 3.8& 0.62\\
                30  GeV& 4.5& 0.5& 4.0& 0.65\\
                40  GeV& 4.7& 0.6& 4.1& 0.67\\
                60  GeV& 5.1& 0.6& 4.5& 0.73\\
\hline
\multicolumn{5}{c}{b) 3 parton calculation}\\
& $B\cdot\sigma(m_H=800~\rm GeV)$& $B\cdot\sigma(m_H=100\rm~GeV)$&
$B\cdot\sigma_{\rm signal}$& $\varepsilon$\\
$p_T^{\rm veto}=20$ GeV& 3.4& 0.2& 3.1& 0.50\\
                30  GeV& 3.6& 0.2& 3.4& 0.54\\
                40  GeV& 3.8& 0.2& 3.6& 0.57\\
                60  GeV& 4.3& 0.3& 3.9& 0.62
\end{tabular}
\end{table}


\begin{figure}\caption{
\label{figone} Representative Feynman graphs contributing to the
electroweak process $pp\to W^+W^-jjjX$. }
\end{figure}

\begin{figure}\caption{
\label{ygap} Gap width distribution, $d\sigma /dy_{gap}$, in the
electroweak process $pp\to W^+W^-jj(j)X$ at $\protect\sqrt{s}=40$~TeV.
Results are shown for the full ${\cal O}(\alpha_s \alpha_{QED}^4)$
calculation and Higgs boson masses of $m_H=800$~GeV (solid line) and
$m_H=100$~GeV (dashed line). The cuts of
Eqs.~(\protect\ref{cutell})--(\protect\ref{defgap}) are imposed, including
lepton acceptance cuts of $p_{T\ell} > 100\; {\rm GeV}$, $|\eta_\ell| < 2$.
The cross section is normalized to include all $W$ decay modes, however. }
\end{figure}

\begin{figure}\caption{
\label{ysclosest} Distribution in $\Delta\eta_{sj}$, the minimum pseudorapidity
separation of the third (typically soft) parton from the two jets defining the
boundary of the gap region. Negative values of $\Delta\eta_{sj}$ correspond to
the emission of a soft parton ($p_T<40$~GeV) into the gap region. Curves are
shown for the same parameters as in Fig.~\protect\ref{ygap}. }
\end{figure}

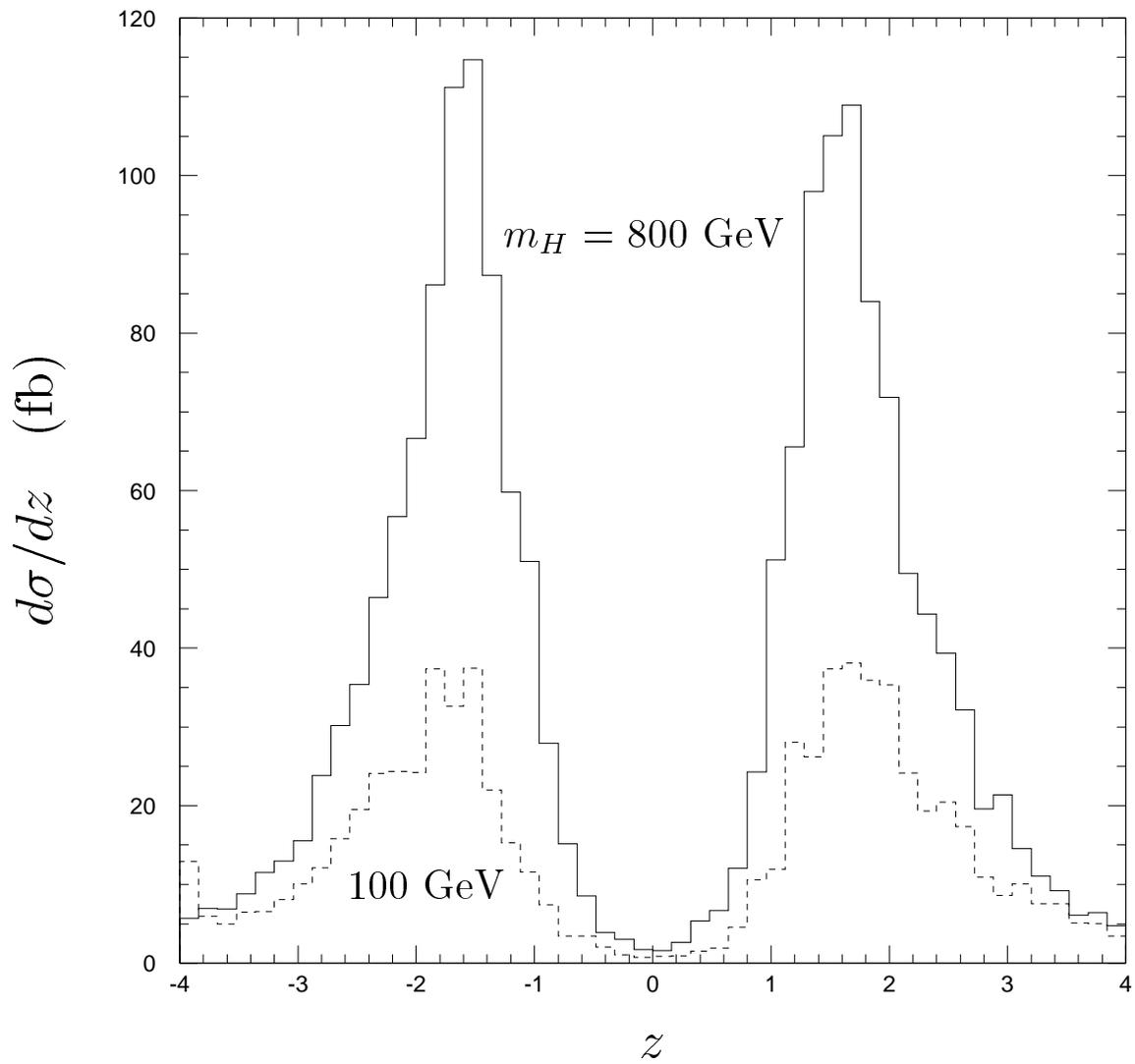
\begin{figure}\caption{
\label{dsigdz} Normalized gap width distribution, $d\sigma/dz$ (see
Eq.~(\protect\ref{defofz})), for
$W^+W^-jj(j)X$ events in $pp$ collisions at $\protect\sqrt{s}=40$~TeV. $z=0$
corresponds to the center of the gap and $z=\pm 1$ mark the gap edges. As in
Fig.~\protect\ref{ygap}
the cross sections are normalized to include all $W$ decay modes. Results are
shown for $m_H=800$~GeV (solid line) and $m_H=100$~GeV (dashed line) and
the cuts of Eqs.~(\protect\ref{cutell})--(\protect\ref{defgap}) are imposed.  }
\end{figure}

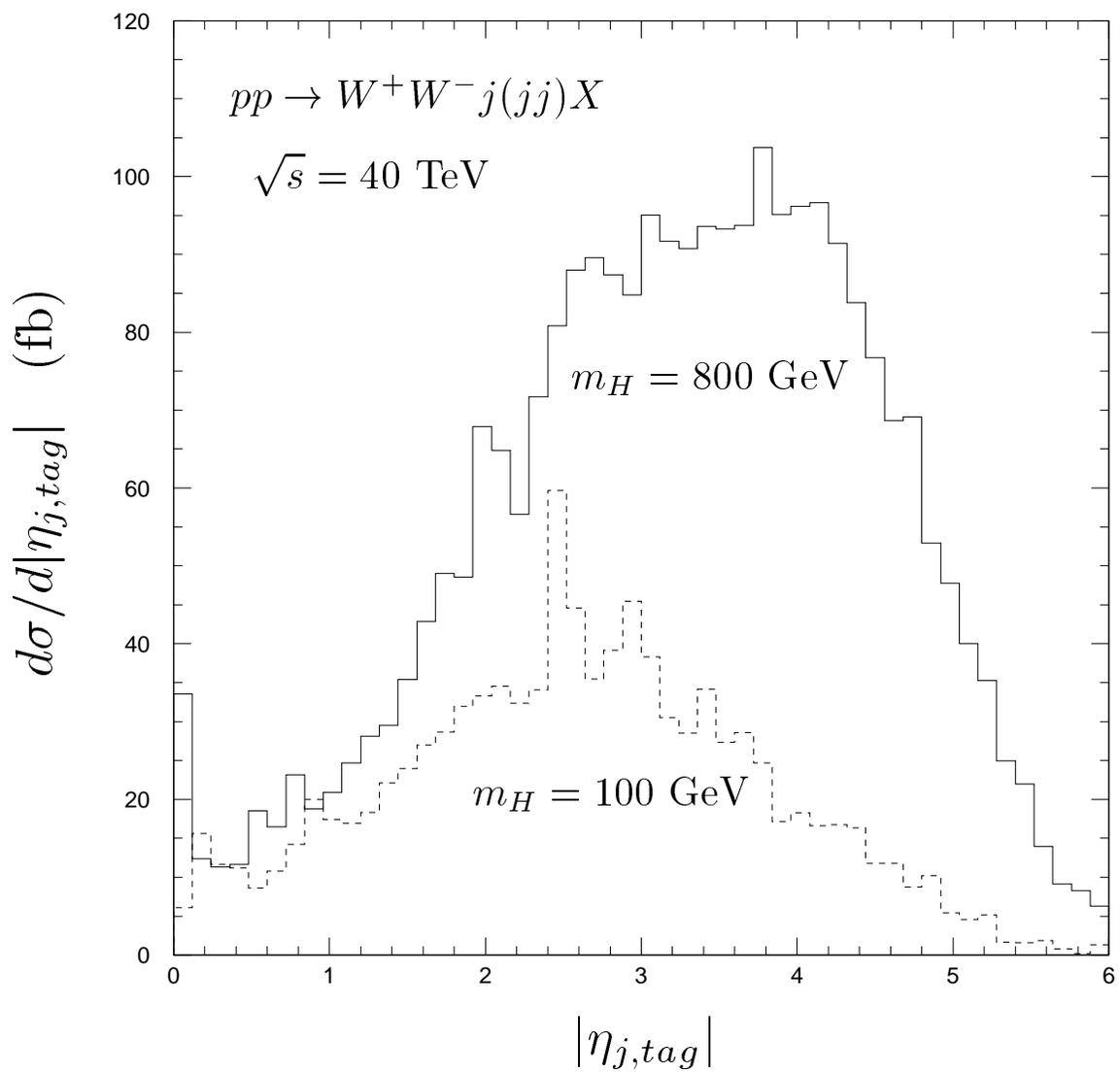
\begin{figure}\caption{
\label{ytag} Pseudorapidity distribution of the tagging jet candidate
(the most energetic jet of $p_{Tj}>40$~GeV) in $pp\to W^+W^-j(jj)X$ events at
$\protect\sqrt{s}=40$~TeV. Curves are shown for 1-jet inclusive events
at ${\cal O}(\alpha_s)$ and Higgs boson
masses of $m_H=800$~GeV (solid line) and $m_H=100$~GeV (dashed line). The
acceptance cuts of Eq.~(\protect\ref{cutell2}) and Eq.~(\protect\ref{ptjettag})
are imposed and the
curves are normalized as in Fig.~\protect\ref{ygap}.  }
\end{figure}

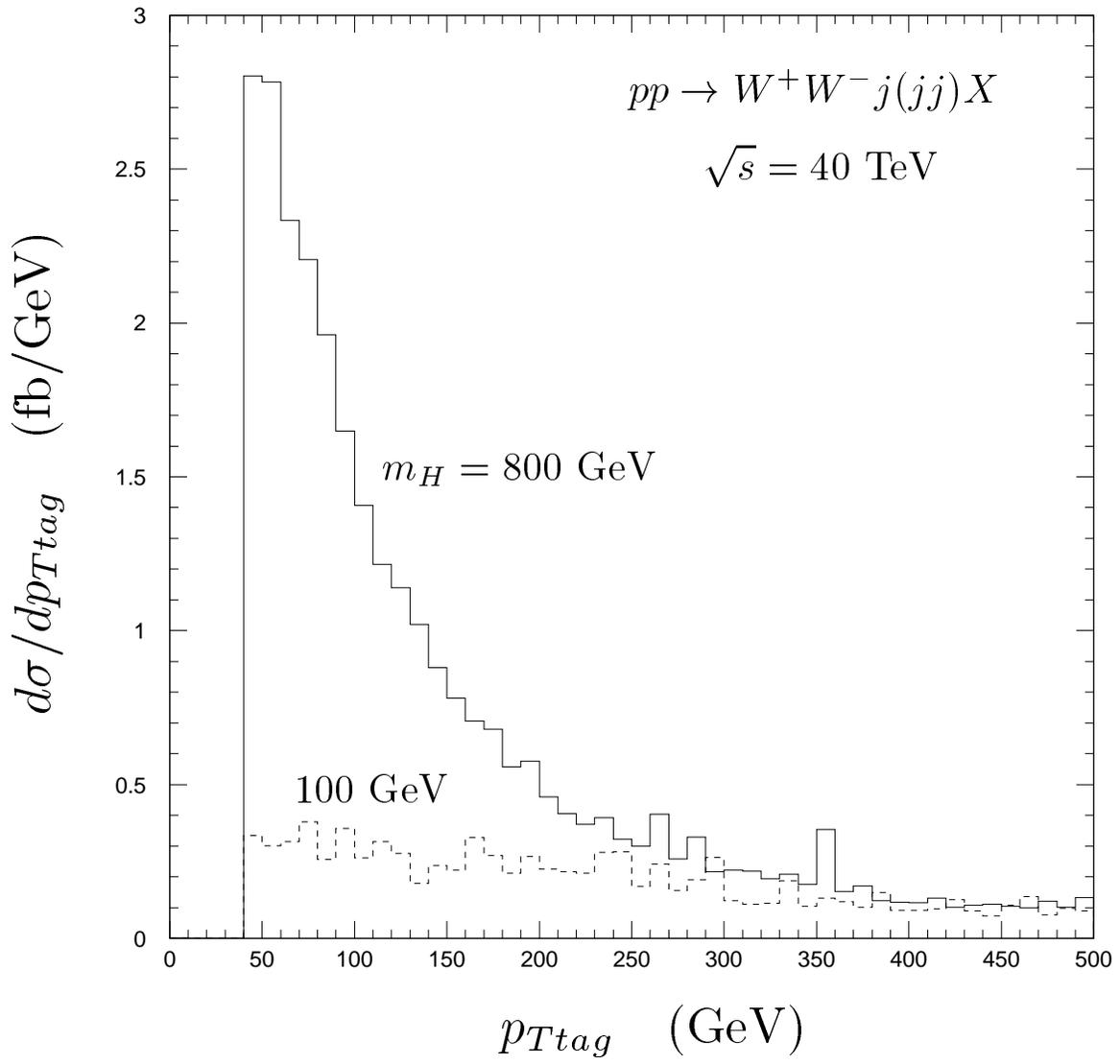
\begin{figure}\caption{
\label{pttag} Transverse momentum distribution of the tagging jet
candidate in $pp\to W^+W^-j(jj)X$ events at $\protect\sqrt{s}=40$~TeV.
Acceptance cuts and normalization of the curves are as in
Fig.~\protect\ref{ytag}. }
\end{figure}

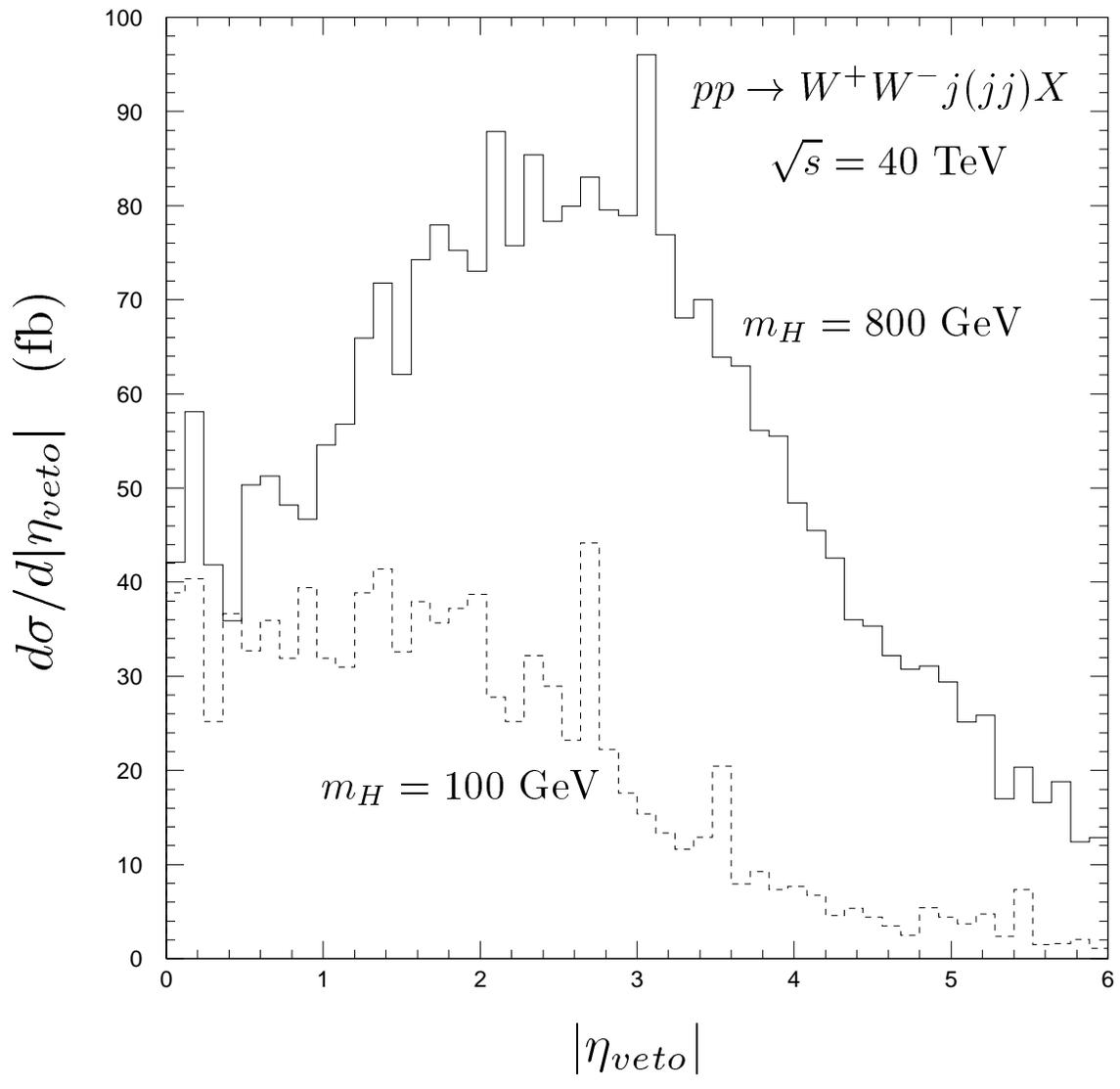
\begin{figure}\caption{
\label{etaveto} Pseudorapidity distribution of the veto-jet candidate
in $pp\to W^+W^-j(jj)X$ events at $\protect\sqrt{s}=40$~TeV. Curves are
shown for 1-jet inclusive events
at ${\cal O}(\alpha_s)$ and Higgs boson masses of $m_H=800$~GeV (solid line)
and $m_H=100$~GeV (dashed line). In addition to the acceptance cuts as imposed
in Figs.~\protect\ref{ytag} and \protect\ref{pttag} the presence of a tagging
jet of $p_{Tj}>40$~GeV, $E_j>1$~TeV is required in the pseudorapidity range
$3<|\eta_{j,\rm tag}|<5$. The curves are normalized as in
Fig.~\protect\ref{ygap}.  }
\end{figure}

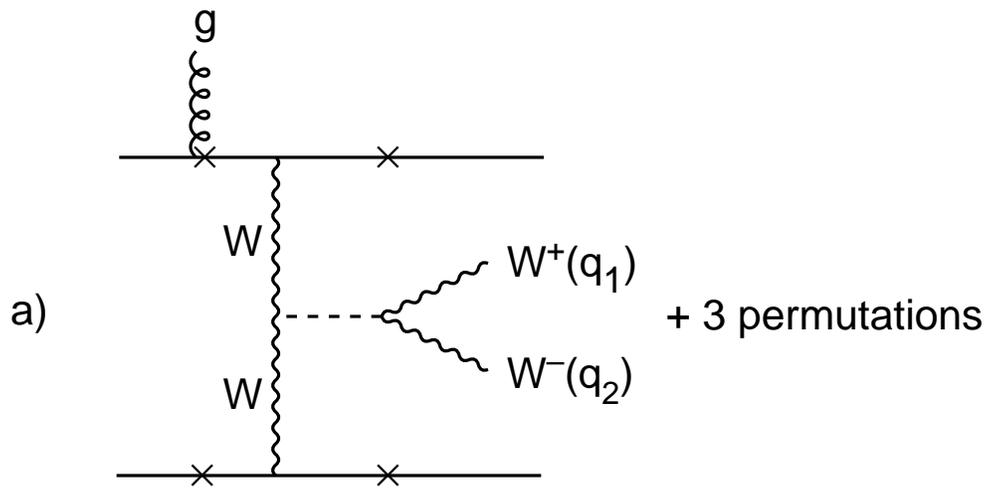
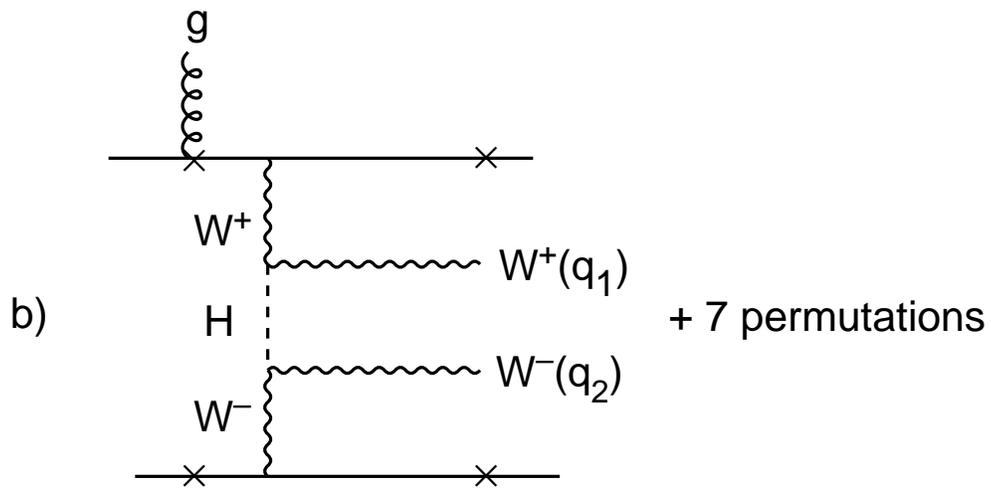
\begin{figure}\caption{
\label{FigFeynAB}
Feynman diagrams corresponding to amplitudes ${\cal M}_1$ through ${\cal M}_3$.
${\cal M}_3$ is obtained from ${\cal M}_2$ by interchanging the $W^+$ and
$W^-$ lines in graph b). The crosses mark the four locations where the
external gluon must be attached, leading to a total of eight permutations for
graph b).           }
\end{figure}

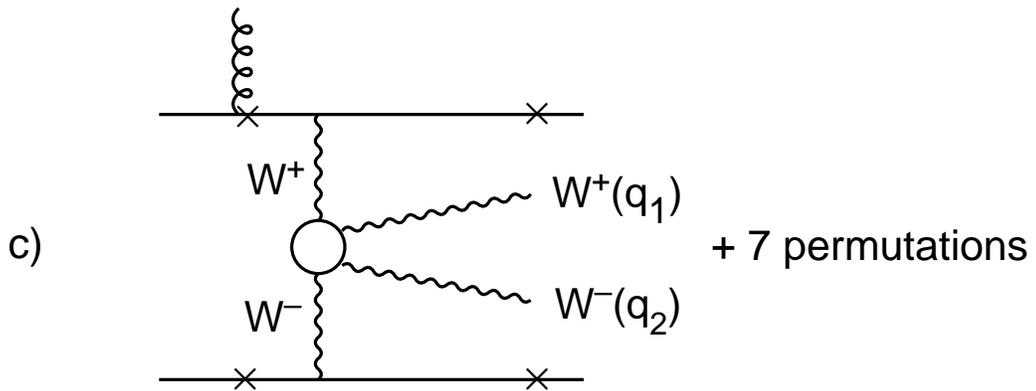
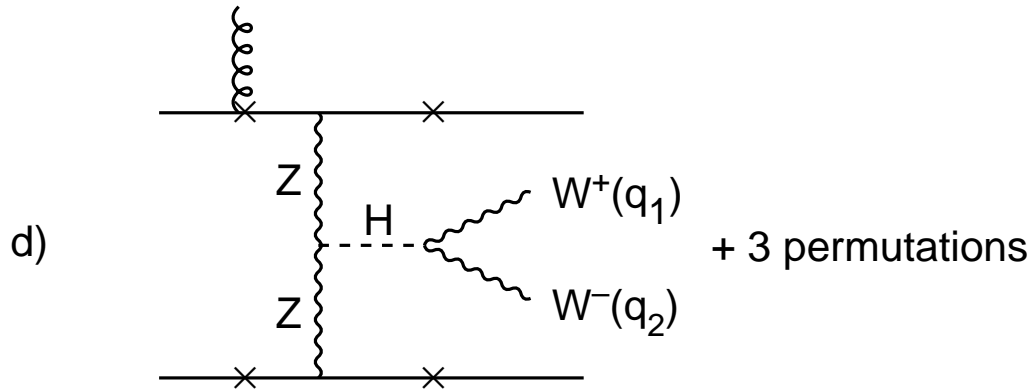
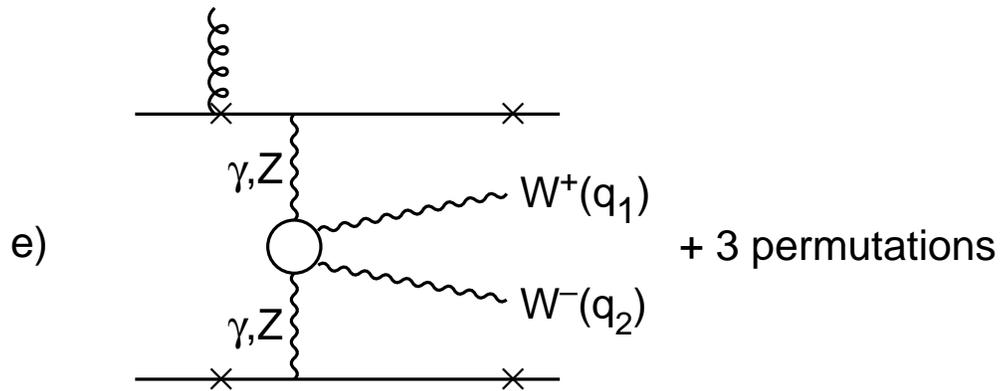
\begin{figure}\caption{
\label{FigFeynCE}
Feynman diagrams corresponding to amplitudes ${\cal M}_4$ through ${\cal M}_7$.
The circles represent the four-boson-vertices as well as $s$-, $t$- and/or
$u$-channel electroweak boson exchange which, in HELAS, are combined into a
single subroutine.   }
\end{figure}

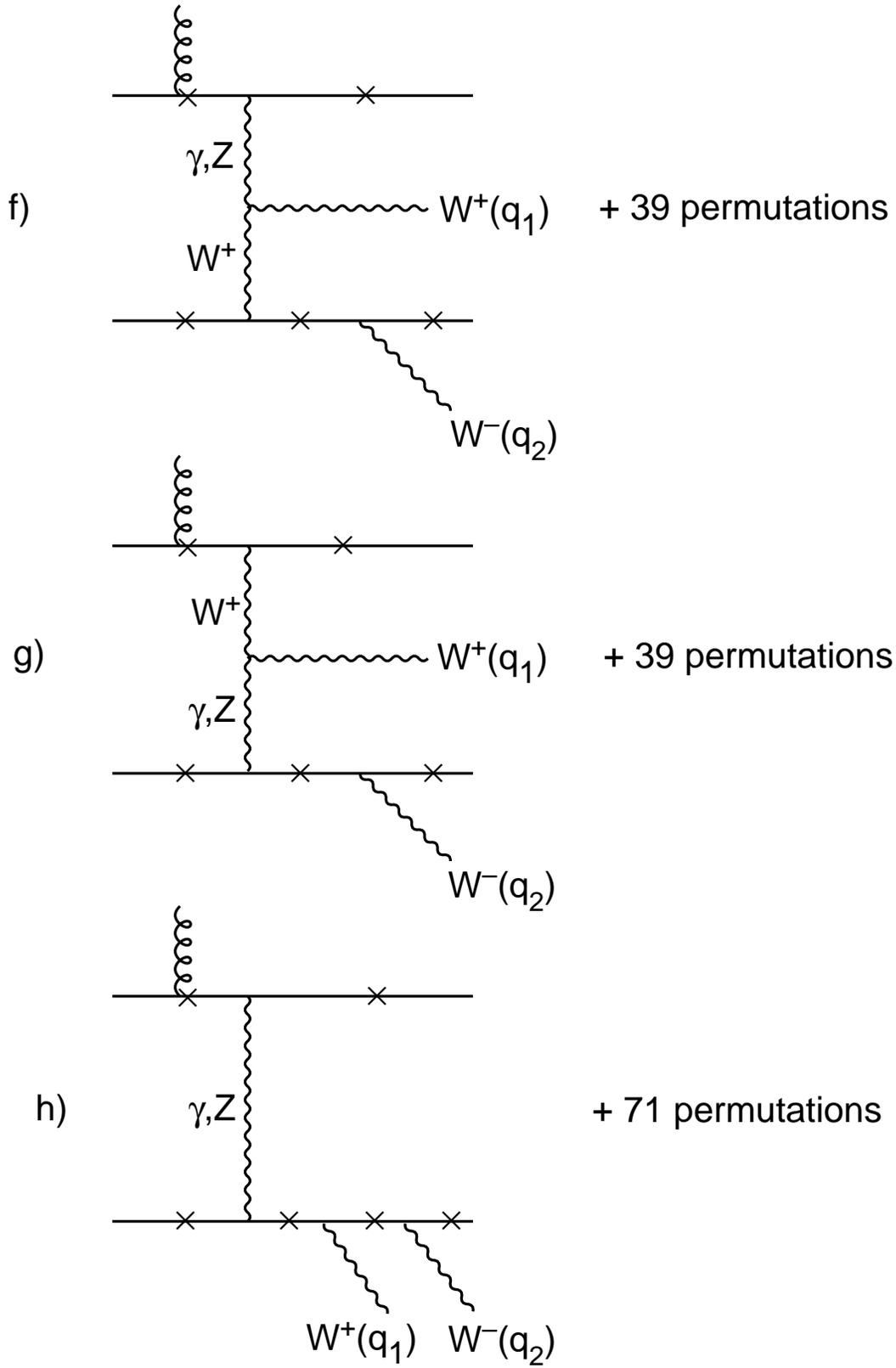
\begin{figure}\caption{
\label{FigFeynFH}
Feynman diagrams corresponding to amplitudes ${\cal M}_8$ through
${\cal M}_{35}$. The permutations involve the different attachments of the
final state $W$'s to the lower quark line, the interchange of final state
$W^+$ and $W^-$, the interchange of upper and lower quark line, and the
different attachments of the gluon.     }
\end{figure}

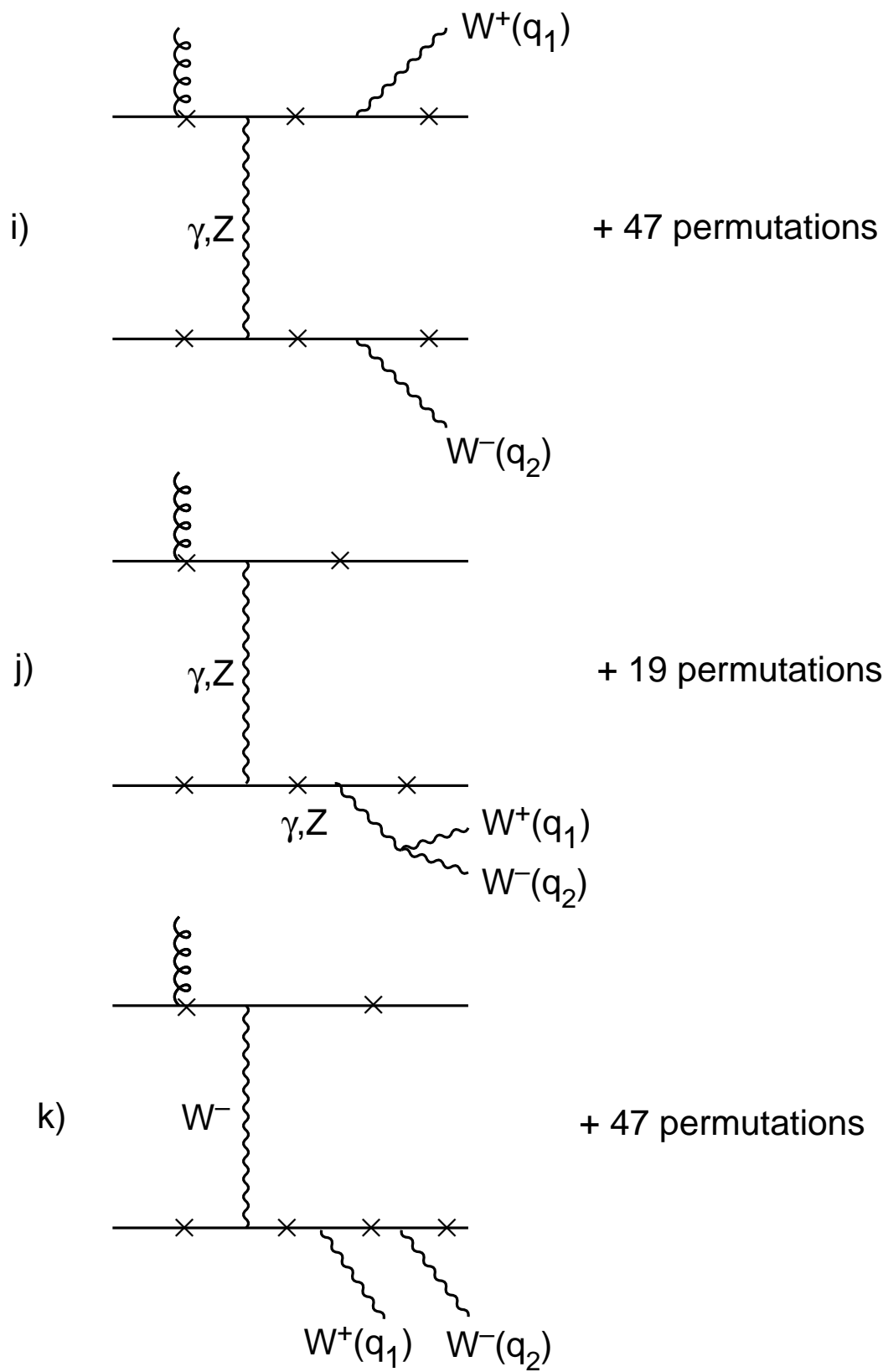
\begin{figure}\caption{
\label{FigFeynIK}
Feynman diagrams corresponding to amplitudes ${\cal M}_{36}$ through
${\cal M}_{55}$. In determining the number of permutations, $\gamma$ and $Z$
exchange are not counted separately.     }
\end{figure}

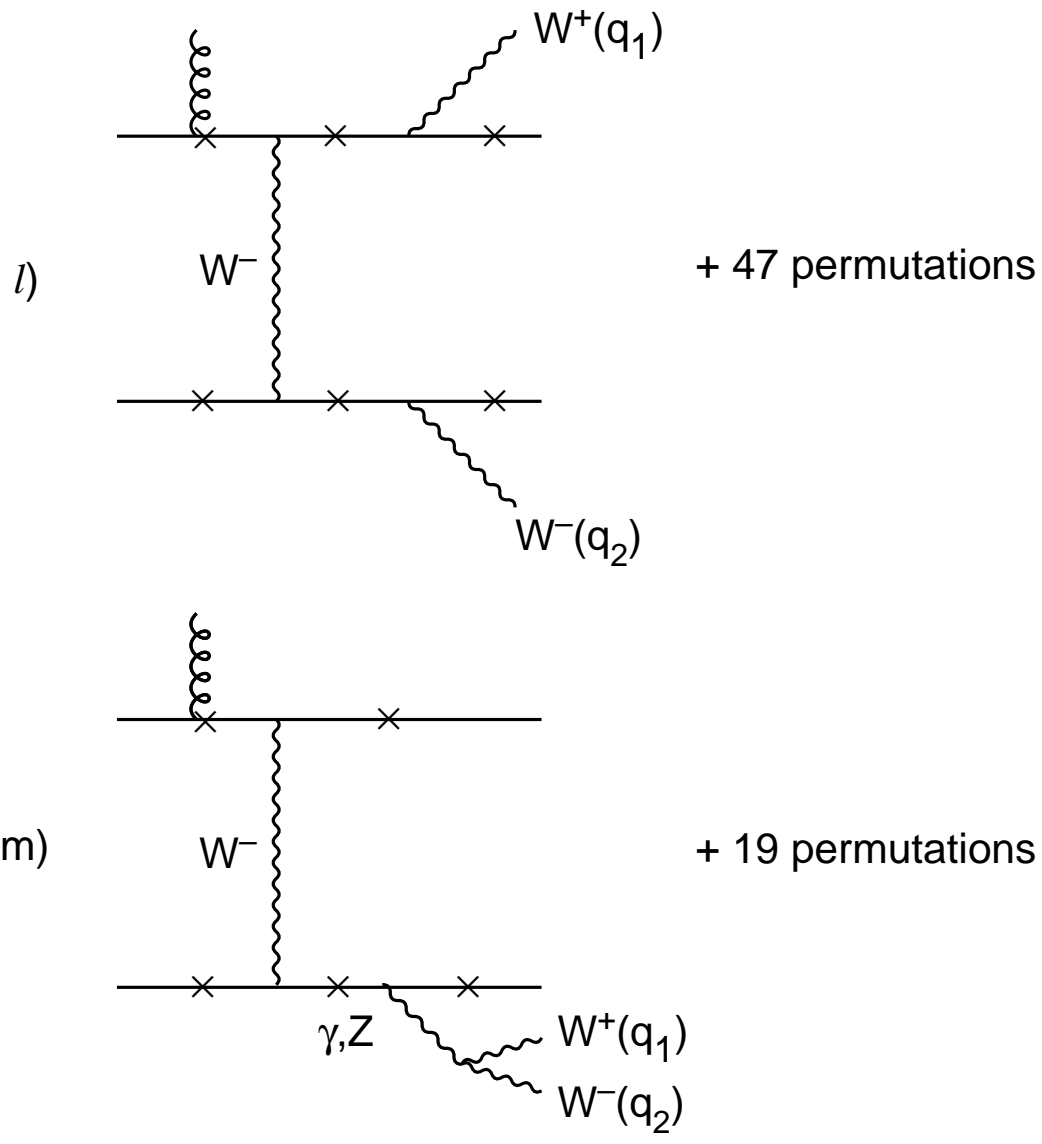
\begin{figure}\caption{
\label{FigFeynLM}\leftline{
Feynman diagrams corresponding to amplitudes ${\cal M}_{56}$ through
${\cal M}_{67}$.}
           }
\end{figure}

\end{document}